\documentclass[10pt,a4]{article}
\usepackage[final]{nips_2017} 
\usepackage[utf8]{inputenc}
\usepackage[dvipsnames]{xcolor}
\usepackage{graphicx} 
\usepackage[tight,footnotesize]{subfigure}
\usepackage{multirow}
\usepackage{url}

\usepackage{enumitem}
\usepackage{amssymb}
\usepackage{amsmath}
\usepackage{mathtools}
\usepackage{pgf} 
\usepackage{tikz} 
\usepackage{filecontents,pgfplots} 
\usepackage{pgfplotstable} 
\usepackage{keyval} 
\usepackage{csquotes}
\usetikzlibrary{calc} 
\usepackage{fp}  
\usetikzlibrary{shadows} 
\usetikzlibrary{arrows,automata,shapes,patterns,backgrounds} 
\usetikzlibrary{positioning,calc,decorations,fit} 
\usepgflibrary{decorations.markings} 
\usepackage[nolist]{acronym} 
\usepackage{etoolbox} 
\usepackage[binary-units=true]{siunitx}

\input{acropatch.tex} 
 
\usepackage{textcomp} 
\pgfplotsset{compat=newest} 
 
\usepackage{subfigure} 
\usepackage{ctable} 
 
\usepackage{prettyref}

\usepackage{todonotes}

\usepackage[ruled,linesnumbered]{algorithm2e}

\usepackage{listofsymbols}

\newenvironment {revision2}
{\color{black}}
{\color{black}}    

 \SetAlFnt{\small}
 \SetAlCapFnt{\small}
 \SetAlCapNameFnt{\small}
 \SetAlCapHSkip{0pt}
 \IncMargin{-\parindent}

\opensymdef
\newsym[application graph]{Gapp}{G_A(V,E)}
\newsym[set of application graph vertices]{Vapp}{V}
\newsym[set of tasks]{tasks}{T}
\newsym[task]{task}{t}
\newsym[deadline]{deadline}{\delta}
\newsym[set of edges]{edges}{E}
\newsym[set of messages]{messages}{M}
\newsym[message]{mess}{m}
\newsym[period of the application]{period}{P}
\newsym[worst case execution time]{wcet}{W}
\newsym[message size]{size}{size}
\newsym[architecture graph]{Garch}{G_{arch}(U,L)}
\newsym[short notation of architecture graph]{Garchs}{G_{arch}}
\newsym[hop distance]{hop}{hops}
\newsym[set of links]{links}{L}
\newsym[ink]{link}{l}
\newsym[number of scheduling intervals on the NoC]{SLmax}{SL_{max}}
\newsym[service level NoC]{SL}{SL}
\newsym[router delay NoC]{rdelay}{\Delta_{Rf}}
\newsym[cycle length]{cycle}{\tau}
\newsym[frequency]{frequency}{f}
\newsym[capacity]{capacity}{cap}
\newsym[bandwidth]{bandwidth}{bw}
\newsym[number of flits]{flits}{flits}
\newsym[set of processing elements]{pes}{U}
\newsym[processing element]{pe}{u}
\newsym[ressource type]{rtype}{rtype}
\newsym[ressource type function]{type}{type}
\newsym[set of ressource types]{RES}{R}
\newsym[ressourcetyp]{res}{r}
\newsym[target pe]{targetpe}{\pe_{1}}
\newsym[target pe]{targetpetwo}{\pe_{2}}

\newsym[power]{power}{power}
\newsym[path of task graph]{tgpath}{X}

\newsym[scheduling interval]{Snt}{SI}
\newsym[scheduling interval os overhead]{SIos}{SI_{os}}
\newsym[maximum number of tasks for schedule]{Kmax}{K_{max}}
\newsym[end to end latency]{EEL}{\mathcal L}
\newsym[Priority]{prio}{pr}
\newsym[predecessor]{pred}{pred}
\newsym[successor]{suc}{suc}

\newsym[constraint graph]{Gconstraint}{G_C(V_C, E_C)}
\newsym[short notation of constraint graph]{Gconstraints}{G_C}
\newsym[constraint graph's vertices]{Vconstraint}{V_C}
\newsym[task clusters]{taskclusters}{T_C}
\newsym[task cluster]{taskcluster}{C}
\newsym[communication channels]{commchannels}{M_C}
\newsym[communication channel]{commchannel}{B}
\newsym[external predecessor function]{predf}{e_{pre}}
 
\newsym[list of priorities]{prios}{\left\langle pr(t),~\forall t \in C\right\rangle}

\newsym[list of predecessors]{preds}{\left\langle e_{pre}(t_0),..,e_{pre}(t_{K-1}) \right\rangle }

\newsym[map task cluster]{map}{\beta}
\newsym[route message cluster]{route}{\rho}

\newsym[map task DSE]{mapDSE}{\beta_{DSE}}
\newsym[route message DSE]{routeDSE}{\rho_{DSE}}

\newsym[map task cluster]{mapCG}{\beta_{CG}}
\newsym[route message cluster]{routeCG}{\rho_{CG}}
\closesymdef

\begin{document}
\begin{center}
	{\Large
		\textbf\newline{A Design-Time/Run-Time Application Mapping Methodology for Predictable Execution Time in MPSoCs}
	}
	\newline
	\\
	Andreas Weichslgartner\textsuperscript{1*},
	Stefan Wildermann\textsuperscript{2},
	Deepak Gangadharan\textsuperscript{3}, \\
	Michael Gla{\ss}\textsuperscript{4},
	J\"urgen Teich\textsuperscript{2}

\end{center}	
\begin{acronym}

\acro{NoC}{network-on-chip}

\acro{SoC}{system-on-chip}

\acro{WCET}{worst-case execution time}

\acro{DSE}{design space exploration}

\acro{VC}{virtual channel}

\acro{NI}{network interface}

\acro{QoS}{quality-of-service}

\acro{CSP}{constraint satisfaction problem}

\acro{MPSoC}{multiprocessor system-on-chip}

\acro{PE}{processing element}

\acro{RM}{run-time management}

\acro{GS}{guaranteed service}

\acro{TDMA}{time division multiple access}

\acro{EA}{evolutionary algorithm}

\acro{ILP}{integer linear programming}

\acro{SA}{simulated annealing}

\acro{OS}{operating system}

\acro{OP}{operating point}

\acro{CDF}{cumulative distribution function}

\acro{HAM}{hybrid application mapping}

\acro{MMKP}{multiple-choice knapsack problem}

\end{acronym}

\begin{abstract}
Executing multiple applications on a single MPSoC brings the major challenge of satisfying multiple quality requirements regarding real-time, energy, etc.
Hybrid application mapping denotes the combination of design-time analysis with run-time application mapping.
In this article, we present such a methodology, which comprises a design space exploration coupled with a formal 
performance analysis.
This results in several resource reservation configurations, optimized for multiple objectives, with verified real-time guarantees for 
each individual application.
The Pareto-optimal configurations are handed over to run-time management which searches for a suitable mapping according to this information. To provide any real-time guarantees, the  performance analysis needs to be composable and the influence of the applications on each other has to be bounded. We achieve this either by spatial or a novel temporal isolation for tasks and by exploiting  composable NoCs.
With the proposed temporal isolation, tasks of different applications can be mapped to the same resource while with spatial isolation, one computing resource can be exclusively used by  only one application. 
The experiments reveal that the success rate in finding feasible application mappings can be increased by the proposed temporal isolation by up to $30\%$ and energy consumption can be reduced compared to spatial isolation.

\end{abstract}

\newcommand\blfootnote[1]{%
	\begingroup
	\renewcommand\thefootnote{}\footnote{#1}%
	\addtocounter{footnote}{-1}%
	\endgroup
}
%
%


\blfootnote{
\textsuperscript{*} Corresponding Author\\
Author's addresses: \textsuperscript{1}Andreas Weichslgartner, Audi Electronics Venture GmbH, Sachsstraße 20, 85080 Gaimersheim, Germany; email: andreas.weichslgartner@audi.de. \textsuperscript{2}Stefan Wildermann, Jürgen Teich
Hardware/Software Co-Design, Department of Computer Science, Friedrich-Alexander-Universität Erlangen-Nürnberg (FAU), Cauerstr. 11, 91058 
Erlangen, Germany; email: \{stefan.wildermann, juergen.teich\}@fau.de.
\textsuperscript{3}Deepak Gangadharan, Department of Computers and Information Science,  University of Pennsylvania, 275 Levine Hall, 3330 Walnut Street, Philadelphia, PA 19104, USA; email: deepakg@seas.upenn.edu.
\textsuperscript{4}Michael Gla{\ss}, Institute of Embedded Systems/Real-Time Systems, Ulm University,
Albert-Einstein-Allee 11, 89081 Ulm, Germany; email:  michael.glass@uni-ulm.de
}

\section{Introduction}
Modern \acp{MPSoC} contain an increasing number of heterogeneous resources, i.e., \acp{PE}, distributed memories, and parallel 
communication interconnects. 
This advances the admittance of more and more functionality into a single chip, which is becoming a prerequisite for implementing modern mobile and multimedia devices, as well as near-future automotive and avionics multi-/many-core systems with varying mixes of concurrently running real-time applications. 
 These application mixes are not always known a priori: At design time, applications may stem from different developer teams and/or are added at different points of time to the already running system. 
Also, the number of possible application mixes is exponential to the number of applications.
This renders the analysis of all application mixes practically infeasible, even if all applications would be known.
In this context, \emph{\acf{RM}} has the purpose of partitioning the system resources and mapping applications onto these partitions dynamically in such a way that certain objectives such as energy consumption are optimized.
For this task, \ac{RM} has to be able to anticipate the impact of the different mapping options on (a) the individual application objectives and (b) the overall system objectives, e.g. resource utilization. 
In recent years, several approaches have emerged tackling the problem via design-time, run-time and combined design- and run-time techniques (termed \emph{hybrid}), see~\cite{Singh:2013a} for an overview.

\begin{figure*}[t]
	\centering
		\includegraphics[width=0.9\linewidth]{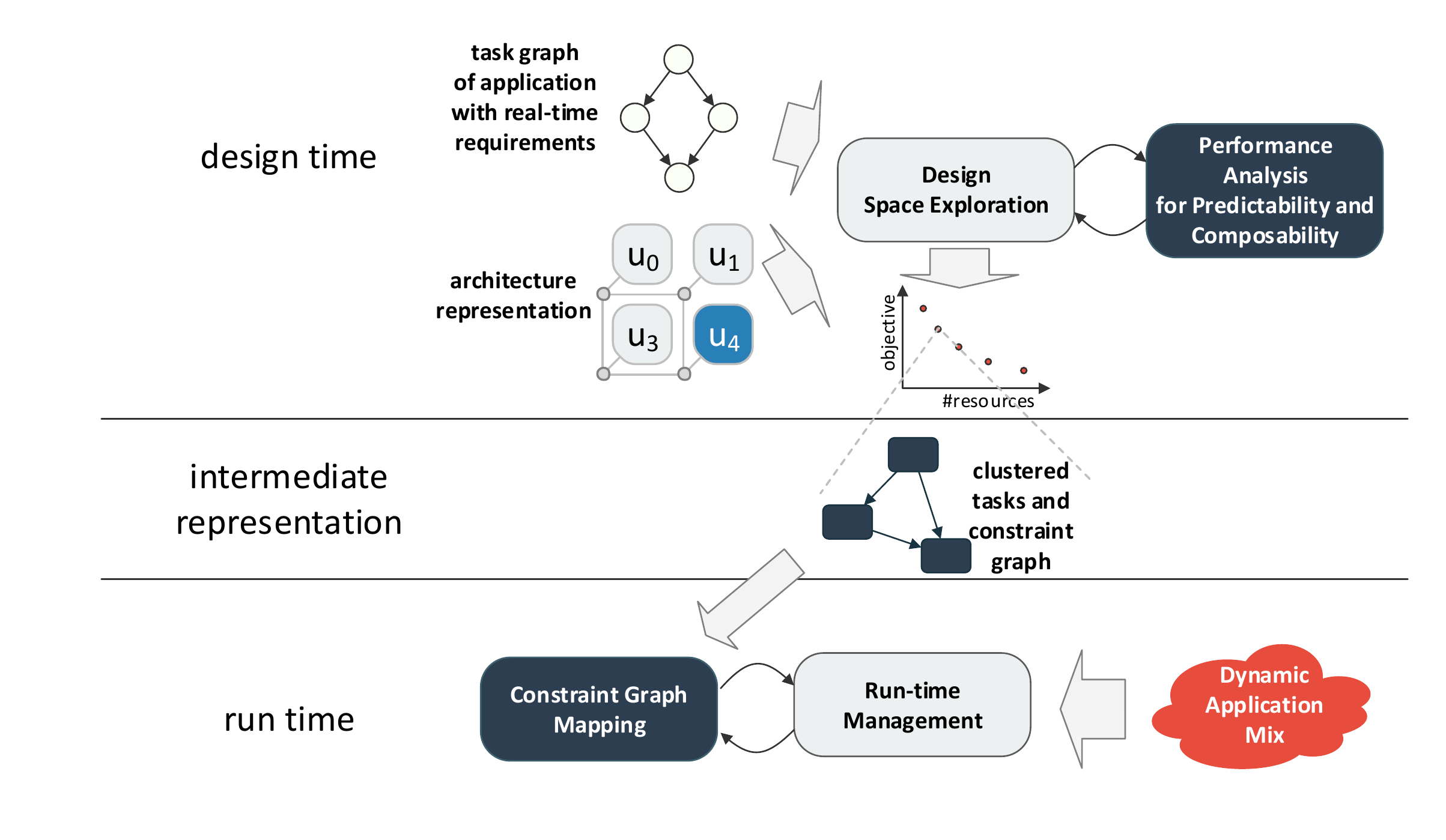} 
	\caption{Overview of the presented hybrid mapping methodology.}
	\label{fig:overview_general}
\end{figure*}

In this article, we present a novel hybrid application mapping methodology outlined in Fig.~\ref{fig:overview_general}. 
The methodology enables the dynamic management of multiple real-time applications with high utilization of available system resources. 
Based on a formal specification of an application by a task graph, different mapping candidates are generated and evaluated with respect to their resource requirements and obtainable execution qualities, called \emph{quality numbers}, at design time. 
To deal with (hard) real-time requirements, a \emph{performance analysis} is proposed to determine worst-case 
latencies, and mapping candidates that do not fulfill deadline constraints are immediately rejected. 
The result is a set of so-called \emph{\acp{OP}} which are non-dominated regarding their resource requirements and quality numbers. 
The idea is that this reduced  information is then used by the \ac{RM} to find a suitable application mapping at run time in a highly predictable fashion, however, with a lower complexity than when having to explore the complete search space without exploration at design time.

However, as soon as multiple individually designed applications are executed on the same system, there would exist side effects in case they share common resources like \acp{PE}, memory, or the communication infrastructure which makes static analysis worthless. 
The main challenge of hybrid application mapping is therefore to guarantee that an application's execution properties, which were analyzed at design time, will actually be satisfied at run time when executed together with other applications. 
This is particularly true for applications with \emph{real-time requirements} that have to meet individual task or application deadlines, where unbounded interference would lead to deadline violations. 
For applying hybrid application mapping in this context, it becomes a prerequisite that the system is \emph{composable} 
\cite{AkessonMHAG11}. 
According to Akesson et al. \emph{composability} is a concept to reduce complexity in real-time \ac{MPSoC} systems.
Composability ensures that the concurrent execution of multiple applications on a common system only has a bounded effect on each application's performance value and it thus ensures that all deadlines can still be met when running any mixture of applications.
This is achieved by analyzing each application individually with resource reservations at design time and ensuring that the required resource reservations are provided at run time in the presence of any arbitrary application mix.

A common approach in related work~\cite{Ykman-Couvreur:2006,Shojaei:2013,WGT:2014} is to try to reduce any side 
effects by 
assigning \acp{PE} exclusively to applications, so that only tasks of the same application may share the same \acp{PE}. 
Yet this way of creating \emph{spatial isolation} is realized for \acp{PE} only.
So, these approaches do not consider that the on-chip communication infrastructure is typically shared to realize flexible memory accesses and data transmissions. 
In fact, \cite{WGWGT14} recently showed that approaches which neglect communication are too optimistic.
This basically means that applications pass admission control and are executed although their deadlines could actually be violated. 
%
%
As a solution~\cite{WGWGT14}~propose a hybrid application mapping approach for \acp{MPSoC} with a composable \ac{NoC} architecture to also bound the interferences in communication. 
Isolation of tasks of different applications is still obtained via spatial isolation by exclusive assignment
of  tasks to \acp{PE} which, however, may result in poor \ac{PE} utilization rates.

As a remedy, we present a novel hybrid mapping approach that is \emph{temporal isolation} of applications on both computational and communication
resources. 
In particular, we propose (a) novel composable scheduling and performance analysis techniques, and (b) a constraint-based run-time mapping approach supported by design-time analysis, which enable to bound the interference effects between applications even if they share the same resources. 
This has the major contribution that system resources can be utilized much better and much more efficiently even under real-time constraints.

\begin{figure*}[t]
	\centering
		\includegraphics[width=1\linewidth]{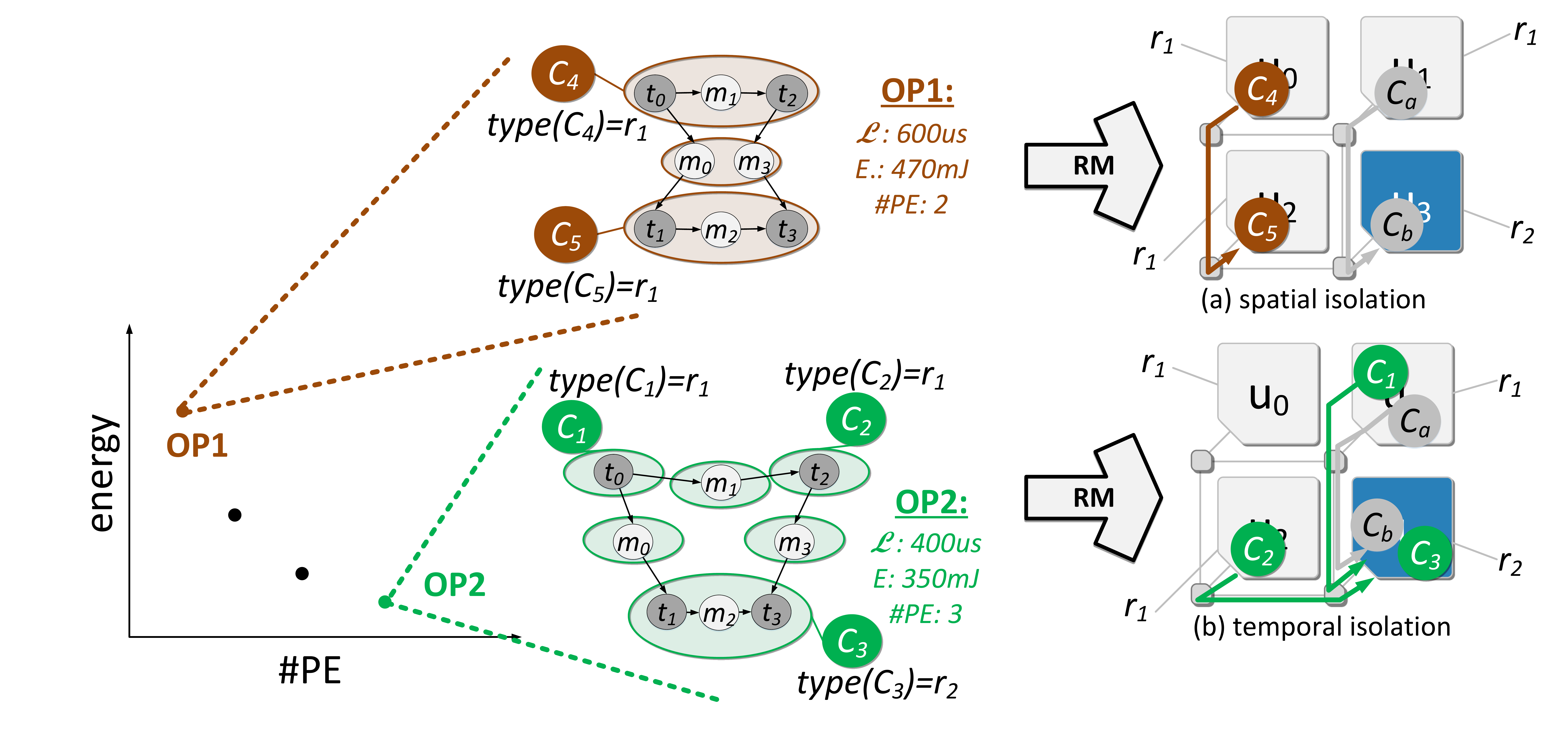} 
	\caption{\label{fig:overview}Schematic overview of spatially and temporally isolated mappings: 
	After \ac{DSE}, the resulting Pareto-optimal \acfp{OP} are stored along with their quality numbers of resources used ($\#PE$), energy consumption ($E$), and their worst-case end-to-end latency ($\EEL$). 
	When the application is released at run time, it needs to be mapped to the system where another application is already admitted
	and currently executed 	(gray clusters $C_a$ and $C_b$). 
	Via spatial isolation, only OP1 can be  feasibly mapped (a) while with our proposed temporal isolation, also OP2 can be mapped (b). 
	This choice results in a lower energy consumption while still meeting the application's deadline by construction. 
	Unused \ac{PE} $u_0$ could be power gated to save more energy.}
\end{figure*}

We illustrate this by means of a \emph{motivational example} according to Fig.~\ref{fig:overview}. 
We assume a given heterogeneous $2 \times 2$ \ac{NoC} target architecture with \acp{PE} being either of resource type $r_1$ or $r_2$. 
An example application, see Fig.~\ref{fig:app}, is specified by a task graph with four tasks $t_i$ and four messages $m_i$. 
Based on this specification, \ac{DSE} is performed (e.g., \cite{BTT:98,LukasiewyczGHT08,dse2parma:2012,WGWGT14}) for generating and evaluating different mappings of tasks to resources. 
By employing static performance analysis, the worst-case end-to-end latencies can be determined for each of these mappings, and mappings that could violate deadlines are rejected. 
The result of the \ac{DSE} is a set of Pareto-optimal \acp{OP} that represents a trade-off between several objectives.
Now, as symmetric architectures may have a huge number of concrete mappings with the same number of \acp{PE} used in the mapping, each \ac{OP} does not describe a concrete mapping of tasks to resources and messages to the \ac{NoC}, but a \emph{constraint graph} instead which describes (a) which tasks are clustered together and (b) mapped onto what resource type to achieve the quality numbers analyzed.
For example, for  \ac{OP}1 in Fig.~\ref{fig:overview}, $t_0$ and  $t_2$ should be mapped together onto a \ac{PE} with resource type  $r_1$ (denoted $C_4$) and $t_1$ and  $t_3$ together onto any available \ac{PE} of resource type  $r_1$ (denoted $C_5$).
For \ac{OP}2 tasks $t_1$ and $t_3$ need to be mapped together onto a \ac{PE} of the type  $r_2$ (denoted $C_3$), while tasks $t_0$ and $t_2$ should be mapped onto two different \acp{PE} of type $r_1$ (denoted $C_1$ and $C_2$).
Overall, OP2 uses more resources (two $r_1$, one $r_2$) than operating point OP1 (two $r_1$, zero $r_2$). 
In this example, task mappings according to operating point OP2 can be executed more efficiently and thus have a lower energy consumption due to the higher degree of parallelism. 
Please thereby note that a constraint graph  stands for a family of concrete and symmetrically identical mappings.
The advantage of this separation of static quality analysis and run-time search for a suitable mapping is to reduce the complexity of run-time mapping to a largest possible extent.

This information is now used by the \ac{RM} prior to starting each application at run time. 
In the example illustrated in Fig.~\ref{fig:overview}, tasks belonging to another application ($C_a$ and $C_b$) are already mapped 
to some resources. 
This means that the \ac{RM} needs to determine a feasible mapping just for the new application. 
Figure~\ref{fig:overview}(a) illustrates a \ac{RM} strategy based on \emph{spatial isolation}. 
Here, the already occupied resources cannot be used for mapping the new application. 
Thus, there does not exist a feasible mapping for operating point OP2, as there is no unoccupied instance of resource type $r_2$ for mapping the tasks represented by $C_3$. 
\Ac{RM} therefore has to test the operating point with next lower energy consumption, i.e., operating point OP1, which then can be feasibly mapped as illustrated in the figure. 
In contrast, the proposed approach is now able to share \acp{PE} under certain conditions as introduced in this article. 
As a consequence, operating point OP2 can be mapped according to Fig.~\ref{fig:overview}(b), resulting in a lower energy consumption, where even the unoccupied \ac{PE} $u_0$ could be powered down. 

As illustrated, the advantage of allowing temporal is to obtain a \emph{higher utilization} of the system resources while 
satisfying predictability requirements on execution time. 
This not only has the direct consequence that a \emph{higher number of applications} can be executed concurrently, 
but it is also possible to execute them on fewer \acp{PE} than when they are reserved exclusively for an application. 
Unused \acp{PE} can be power gated, which may additionally \emph{reduce energy consumption}. 
Moreover, in emerging many-core systems, temporary or even permanent unavailability of hardware resources is expected to be experienced more often 
because of
hardware faults (manufacturing variability and aging) or temperature/power management (cf. \cite{Henkel:2013}).
In this context, the proposed mapping approach \emph{enhances robustness} as it is possible to react to unavailability of 
\acp{PE} by re-mapping affected applications onto the remaining \acp{PE}, which can be shared with other applications.
\begin{revision2}
Overall, the contributions of this article are the following:
\begin{itemize}
	\item This article presents the \acl{HAM} methodology, first introduced in~\cite{WGWGT14}, in more detail and with more examples. 
	This approach combines the strengths of design time, e.g., analysis and compute-intensive optimization, with the flexibility of run-time decision making to cope with dynamism.
	While related work often neglects or simplifies the \ac{NoC} communication, with this \ac{HAM} methodology, timing guarantees for state-of-the-art packet-switched \ac{NoC} architectures can be given.
	\item We enhance this methodology by including the concept of temporal isolation. This, opposed to~\cite{WGWGT14}, enables the sharing of \acp{PE} among different applications while still preserving real-time requirements.
	In consequence, this increases the utilization of the system and enables possibilities for energy saving.
	\item We evaluate execution times of the \ac{RM} through a simulation of embedded hardware which is not considered in~\cite{WGWGT14}. For bounding the execution times, we propose to use the backtracking algorithm with timeout mechanism and outline the implications in the conducted experiment.
\end{itemize}
\end{revision2}

%

The remainder of the paper is outlined as follows:
In Section~\ref{sec:relatedwork}, we give an overview of related work.
We formalize the used model of applications and system architecture in Section~\ref{sec:preliminaries}.
Section~\ref{sec:predictability} describes our formal design-time analysis while Section~\ref{sec:dse} details the  
 design-time optimizations.
In contrast, Section~\ref{sec:runtime} deals with run-time mapping.
In Section~\ref{sec:experiments}, we evaluate our approach through several experiments and conclude our work in
Section~\ref{sec:conclusion}.

\section{Related Work}
\label{sec:relatedwork}
According to \cite{Singh:2013a}, application mapping approaches for embedded multi-/many-cores can be classified as design-time mapping, 
(on-the-fly) run-time mapping, and hybrid (design-time analysis and then run-time use) mapping.
In the following, we give a brief overview of the existing mapping approaches:

\emph{Design-time mapping} approaches require a global view of the system for which application mapping is then optimized. 
While these approaches enable application execution with high predictability, support of varying sets of executed applications and/or unpredictable dynamic workload scenarios is not in their focus. 
In general, there are not any strict requirements on the execution-time of design-time approaches and  they can utilize well-known optimization techniques such as \ac{ILP}~\cite{weija:2010}, \ac{EA}~\cite{Choi:2012}, \ac{SA}~\cite{Orsila:2007}, or  divide-and-conquer~\cite{KangYSBHT12}.

\emph{Run-time mapping} approaches use scalable run-time heuristics to determine application mapping whenever the workload scenario of the system is dynamically changing. 
However, they do neglect or cannot guarantee the predictable execution of applications with (typically hard/soft) real-time requirements.
In contrast to design-time mapping, the execution time and available power for determining a mapping is limited.
In consequence, simple and fast heuristics such as simple nearest neighbor algorithms have been proposed 
here (e.\,g.~\cite{Carvalho:2007,Weichslgartner:2011}). 
{
The objectives for run-time optimization are typically soft real time (e.\,g.~\cite{bbw08}), energy (e.\,g.~\cite{Chou:2008,Holzenspies:2008}), or average speedup (e.\,g.~\cite{Kobbe:2011}).
In \cite{Holzenspies:2008}, an iterative online application mapping methodology for heterogeneous \ac{NoC} architectures is proposed. 
After an initial greedy task to resource  assignment, the mapping is optimized and afterwards it is checked if all \ac{QoS} are met.
If not, the mapping is marked as infeasible and feedback to the previous steps to remap the application is given.
In contrast to this work, we propose to pre-define already mapping classes which define the implementation, i.e. task variant for a certain resource type,   at design time.
}

\begin{revision2}
\emph{\Acf{HAM}} attempts to combine the strengths of design-time and run-time mapping.\end{revision2}
Here, scenario-based (e.\,g.~\cite{Stralen:2010}) and multi-mode (e.\,g.~\cite{Wildermann:2011}) embedded system design tries to optimize the mappings for different workload scenarios or execution modes at design time and then just applies them at run time. 
Yet, considering all possible combinations of applications in different scenarios, of course, would result in a lot of mappings that need to be stored, as the number of combinations increases exponentially with the number of applications.
To reduce this number of  mappings, the authors in~\cite{Wei:15} propose to save only a \enquote{representative subset of scenarios for each cluster}.
For each application, two operating points (throughput-optimized and throughput under a certain energy budget) are stored after \ac{DSE}.
The \ac{RM} then tries to detect a scenario at run time and to customize and optimize the mapping accordingly.
In contrast to this approach, we exploit the concept of composability to explore several mappings per application which can be embedded at run time with guaranteed upper bounds for end-to-end-latency and without the need of scenarios and any run-time optimization.
\par In~\cite{Singh:2013b}, a hybrid mapping methodology that determines energy and throughput optimized application mappings is proposed.
Pareto-optimal mappings with iteratively increased hop distances between the tasks are generated at design time.
At run time, a heuristic selects a mapping based on the number of used processor tiles while only considering the maximal number of hops for the respective operating point. 
This approach is only viable when using a communication infrastructure which provides dedicated point-to-point connections between all pairs of computational resources. 
This has the major advantage that the usage of such end-to-end connections results in fixed communication latencies between computational resources, and thus supports the verification of real-time guarantees. 
However, implementing dedicated connections between all pairs of computational resources is not practicable and scalable in many-core systems with tens or even hundreds of \acp{PE}.

\begin{revision2}
In~\cite{Mariani:10,Ykman-CouvreurAMPSZ11}, \ac{HAM} approaches where a design-time \ac{DSE} generates operating points which are mapped onto a bus-based \acp{MPSoC} during run-time by a light-weight \ac{MMKP} solver are presented.
Another approach for bus-based \acp{MPSoC} which solves the \ac{MMKP} heuristically during run time by using Pareto-Algebra  is presented in~\cite{Shojaei09}.
\cite{JungLKKOH14} propose to explore Pareto-optimal schedules for data-flow modeled applications while a greedy run-time manager performs allocation and binding.
As communication infrastructure they assume a \ac{NoC} \enquote{point-to-point connections with fixed
latency between tiles} and the real-time properties are assured by spatial isolation, i.e., exclusive tile usage by one application. 
\end{revision2}

In fact, sophisticated \ac{NoC} architectures multiplex multiple communication flows over shared resources, i.e., \emph{links}~\cite{dally:2001}.
They perform packet-switched routing by partitioning each communication into packets which are then routed over shared links. 
While this enhances scalability, it makes it harder to give any guarantees regarding the communication latency as this requires a communication infrastructure with \ac{QoS} guarantees.
In order  to give any \ac{QoS} guarantee, each flow can only get a limited time budget of a multiplexing interval.
There are different strategies to assign such budgets, e.g. priority-based~\cite{Carara:2011}, global \ac{TDMA}~\cite{Goossens:2005}, or weighted round robin~\cite{Heisswolf20132603}.

\section{Preliminaries}
\label{sec:preliminaries}
 In the following, we introduce the required formal notations and models for applications as well as the \ac{MPSoC} system architecture. 
 
 \subsection{Application Model}
\begin{figure}[t]
 \centering
\subfigure[task graph]{
\includegraphics[width = 0.26\columnwidth]{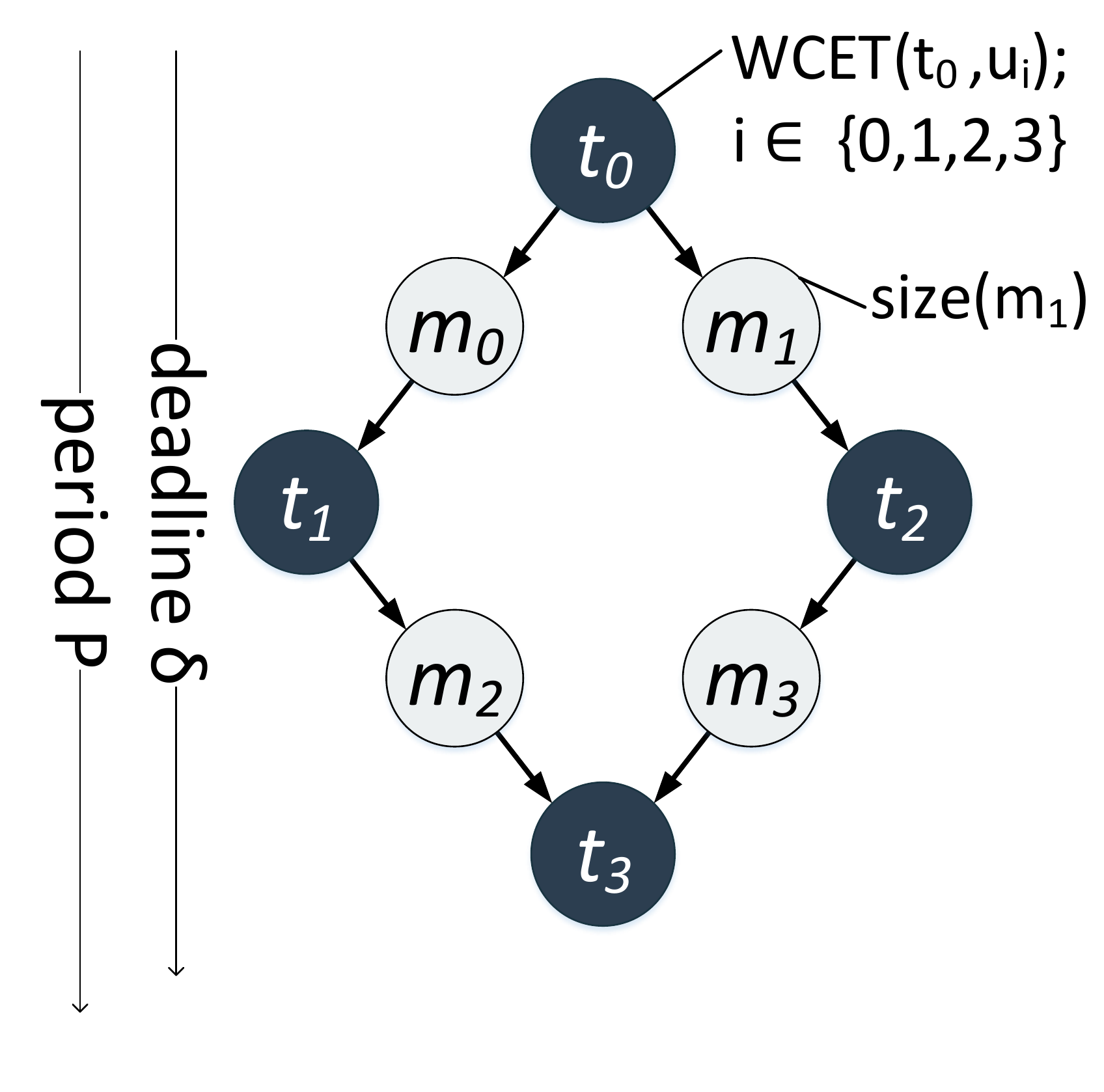}
 \label{fig:app}
}
 \subfigure[system architecture]{
 \includegraphics[width = 0.41\columnwidth]{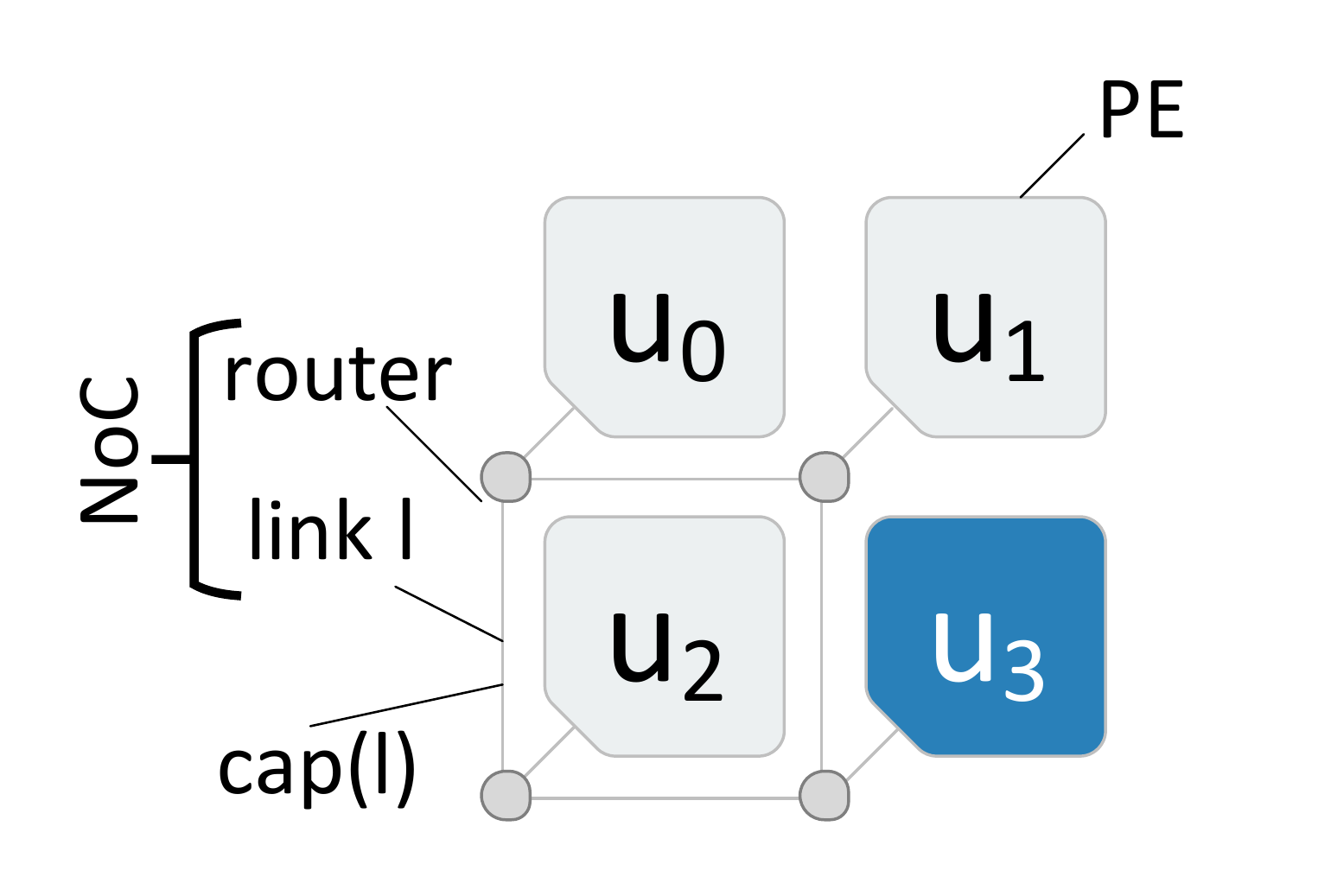}	
 \label{fig:arch}
}
\subfigure[application mapping]{
 \includegraphics[width=0.24\textwidth]{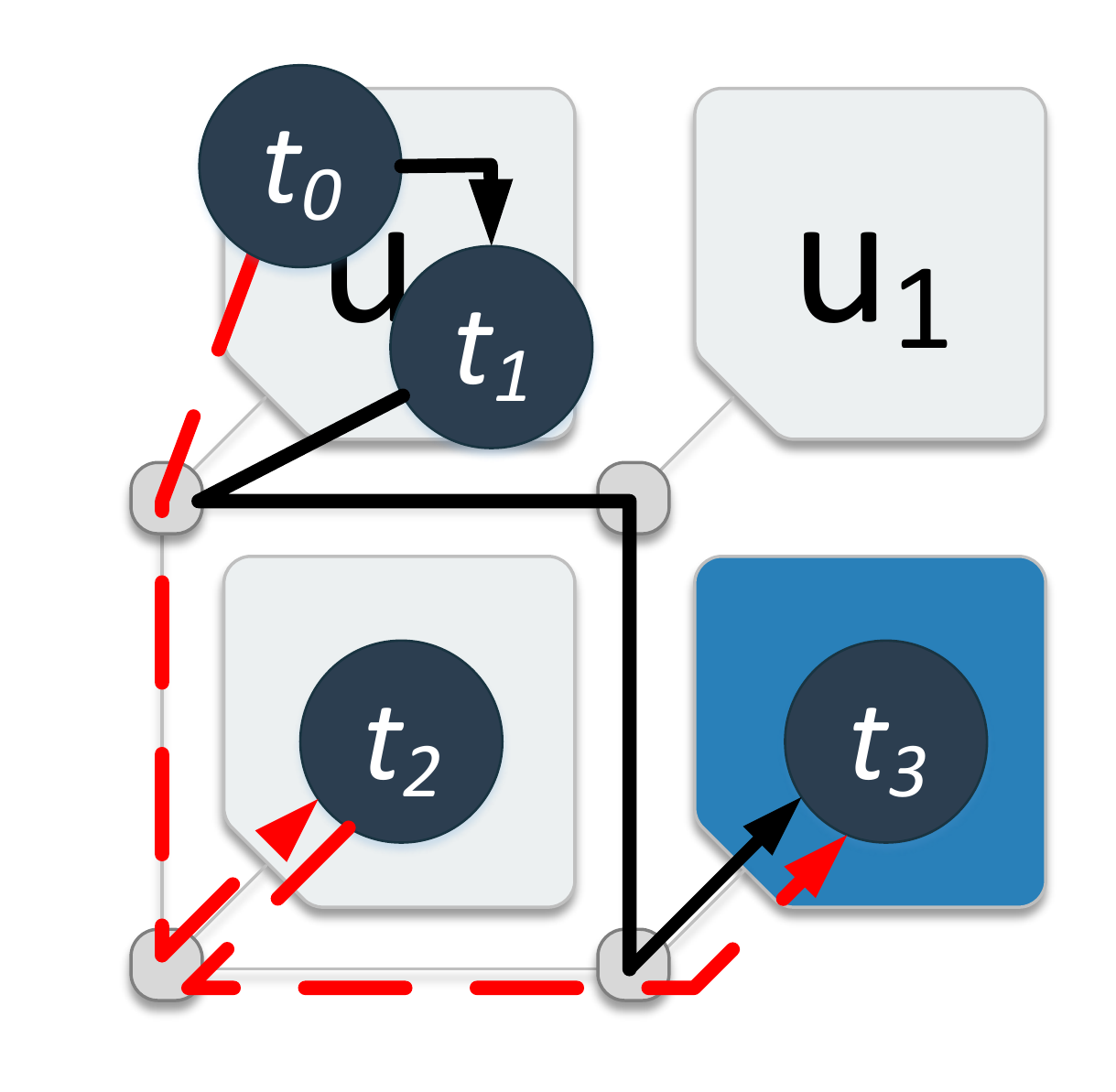}
 \label{fig:example_longest_path}
}
  \caption{Representation of (a) an example application by a task graph and (b) an example system architecture including 4 \acp{PE} with two
different resource types (colored white and gray) and a $2\times 2$ \ac{NoC}.
One possible application mapping of the task graph onto the architecture is shown in (c), also illustrating the two paths in the task graph which are relevant for the calculation of the worst-case end-to-end latency by a solid blue and a dashed red line.
}
\end{figure}

In this work, we concentrate on periodic real-time applications (e.g. image/signal processing, control loops, streaming and multimedia applications, etc.). 
Such applications typically can be represented by a\-cyclic, directed, bipartite \emph{task graphs}. 
Fig.~\ref{fig:app} illustrates an example. 
A task graph is denoted by $\Gapp$. 
The vertices $\Vapp = \tasks \cup \messages$ are composed of the set of \emph{tasks} $\task \in \tasks$, representing sequential code segment, and the set of \emph{messages} $\mess \in \messages$, representing data exchanged between pairs of tasks. 
Consequently, tasks in $\tasks$ are connected through directed edges in $\edges$ with messages  in $\messages$ and vice versa, i.e., $\edges = \tasks \times \messages \cup \messages \times \tasks$.

Applications represented by task graphs shall be executed periodically once admitted with \emph{period} $\period$ and have to meet a certain \emph{deadline} 
$\deadline$.
Furthermore, we assume that the period is at least as long as the deadline.
Every message has a maximum data size $\size(\mess)$ (i.e., payload), so that together with the period, a minimum bandwidth requirement $\bandwidth(\mess)$ can be calculated.
Each task is assumed to represent a sequentially executed code segments of each application, a \emph{\ac{WCET}} $\wcet(\task,\pe)$ can be determined through \ac{WCET} analysis\footnote{We assume architectures with tiles consisting of a single \ac{PE} and analyzable cache, e.g. PowerPC with {partitioned} LRU cache \cite{Wilhelm:2008}. }  for each task $\task \in \tasks$ on resource $\pe$.
The determination of task \acp{WCET} itself is not in the focus of this work but can be derived by \ac{WCET} analysis tools like  aiT~\cite{absinth} or Chronos~\cite{li2007chronos}.
{
However, to prevent cache interferences in private \ac{PE} caches when mapping different tasks to the same \ac{PE}, cache partitioning, private scratchpads, or flushing caches after each scheduling interval (c.f.~\cite{Wilhelm:2008}) may be considered.
}

\subsection{System Architecture} 
 \label{sec:architecture} 
The system architectures $\Garch$ targeted by our approach are many-core systems which
consist of a set of heterogeneous \acp{PE} $\pe \in \pes$.
The resource type $\res \in \RES$ of a \ac{PE} $\pe$ is specified by the function $\type: \pes \rightarrow  \RES$. 
A \ac{NoC} is used as communication infrastructure where routers are connected with each other and to the \acp{PE} via links $\link \in 
\links$ to form a 2-dimensional mesh topology, as exemplified in Fig.~\ref{fig:arch}. 
Each link has a capacity $\capacity(l)$ which is proportional to the link width and the frequency $\frac{1}{\cycle}$.
Moreover, we concentrate on packet-based 
routing, where messages are partitioned into \emph{flits} which are transmitted one after the other over the infrastructure. 
Transmission happens from the sending to the receiving \ac{PE} along a route of consecutive routers. 
The distance between two \acp{PE} $\pe_1$ and $\pe_2$ is determined by a \emph{hop count function} $\hop(\pe_1,\pe_2)$, i.e., the number of routers along the route.

\section{Application Mapping and Static Performance Analysis}
\label{sec:predictability}
The worst-case end-to-end latency of an application depends on its mapping onto the available computing and communication resources.
Using the proposed model, this is formulated as a mapping of the application graph $\Gapp$ onto the architecture graph $\Garch$ obtained by binding each task and routing each message:
\begin{itemize}
 \item[a)] \emph{Binding} $\map{:}~\tasks~{\rightarrow}~\pes$ represents the assignment of each task $\task \in \tasks$ to a target \ac{PE} $\map(\task) \in \pes$.
 \item[b)] \emph{Routing} $\route: \messages \rightarrow 2^{\links}$ represents the routing of each message $\mess$ with sender $\task_1$ and receiver $\task_2$ over a set of connected links $\links^\prime \subseteq \links$ that establish a path between \ac{PE} $\map(\task_1)$ with \ac{PE} $\map(\task_2)$.
\end{itemize}
An example mapping of the introduced task graph $\Gapp$ from Fig.~\ref{fig:app} is shown in Fig.~\ref{fig:example_longest_path}.

For a mapping to be feasible, it must be guaranteed that the end-to-end latency for executing an application does not exceed its deadline $\deadline$. 
The worst-case end-to-end latency of the application depends on the critical path of the mapped task graph. 
For determining the critical path, we calculate the end-to-end latency for each path of $\Gapp$ by summing up the \emph{worst-case execution latencies} $TL$ of all tasks in the path and the \emph{worst-case communication latencies} $CL$ of all messages in the path. 
The worst-case end-to-end latency of a path $path$ for a given binding $\map$ and routing $\route$ may then be calculated according to
\begin{equation}
L(path, \map, \route) = \sum_{\forall t\in path \cap \tasks} TL(t, \map(t)) + \sum_{\forall m\in path \cap \messages} CL(m, \route(m))
.
\label{eqn:pathlat}
\end{equation}

The \emph{worst-case end-to-end latency} is then the latency of the path with the highest worst-case end-to-end latency (i.e., the critical path):
\begin{equation}
\EEL(\map, \route) = \max_{\forall path \in paths(\Gapp)} \{L(path, \map, \route) \}
.
\label{eqn:wce2elatency}
\end{equation}

Figure~\ref{fig:example_longest_path} presents an example where $\Gapp$ basically includes two paths from the source task $t_0$ to the sink task $t_3$.
One path is $(t_0, m_0, t_1, m_2, t_3)$, and the other path is $(t_0, m_1, t_2, m_3, t_3)$. 
In the given mapping, $t_0$ and $t_1$ are mapped together on one \ac{PE} so that $m_0$ does not have to be routed over the \ac{NoC} but can be established by local memory. 
Note that the resulting delay for doing so has to be already included in the \ac{WCET} analysis. 

When permitting to execute the application on resources that are potentially shared with other applications, they may interfere and affect each other's timing behavior. 
For being able to bound this interference, and thus being able to calculate $TL$ and $CL$ without knowing whether and how other applications share resources, composability is required. 
In the following, we describe techniques for composable communication scheduling and composable task scheduling and their respective worst-case analysis as used in this work. 
Both techniques are based on the idea of reserving periodically available time slots for data transmission and task scheduling, respectively. 
\emph{The interesting aspect is that the worst-case execution and communication latencies obtained here can be composable even during run-time
mapping of new tasks into the system if just certain mapping constraints are satisfied}.
This will be explained in detail in Sect.~\ref{sec:runtime}.

\subsection{Composable Communication Scheduling}
In order to provide the desired composability, the \ac{NoC} architecture has to fulfill certain criteria and has to show a predictable
timing behavior.
One \ac{NoC} architecture which adheres to these requirements is proposed in~\cite{Heisswolf20132603}.
 This architecture uses wormhole switching and the concept of \acp{VC} to ensure a high throughput and low latencies.
Further, \ac{GS} connections  supporting \ac{QoS} can be set up and physical links are arbitrated in a weighted round robin fashion for transmitting the flits of the different messages routed over it. 

A number of $\SLmax$ time slots (one slot for transmitting one flit) is periodically available for the overall transmission out of which a budget of $SL(\mess) \le \SLmax$ time slots can be reserved for the transmission of a message $\mess$. 
%
Note that, in contrast to a global synchronous \emph{\ac{TDMA}} like presented in \cite{Goossens:2005}, only the number and not the position of the allocated time slot is fixed.
This increases the utilization while still allowing to compute upper bounds for throughput and worst-case latency.

The worst-case communication latency $CL(\mess, \route(\mess))$ for transmitting message $\mess \in \messages$ depends on the number of flits $\flits(\mess)$, the length of the route $\route(\mess)$, and the number of reserved time slots $\SL(\mess)$ and can be calculated as follows~\cite{Heisswolf20132603}:
\begin{revision2}
\begin{subequations}
 \label{eqn:nocdel}
 \begin{align}
 CL(\mess, \route(\mess)) & = (\flits(\mess) \cdot \frequency^{-1} + \hop\left(\route(\mess)\right) \cdot \rdelay) \label{eqn:nocdel:free}\\
      & + \left(\left\lceil\frac{\flits(\mess)}{\SL(\mess)} \right\rceil - 1 + \hop\left(\route(\mess)\right)\right) \cdot\left(\SLmax-\SL(\mess)\right) \cdot \frequency^{-1}  .\label{eqn:nocdel:blocked}  
 \end{align}
\end{subequations}
\end{revision2}
\begin{revision2}
In Eq.~\ref{eqn:nocdel:free}, $\rdelay$ denotes the delay for routing one flit in one router with the frequency $\frequency$.
Once the routing decision has been made in one router, one flit per clock cycle ($\frequency^{-1}$) can be transmitted.
\end{revision2}

\begin{figure*}[t]
\subfigure[Best Case]{
\includegraphics[width=0.45\columnwidth]{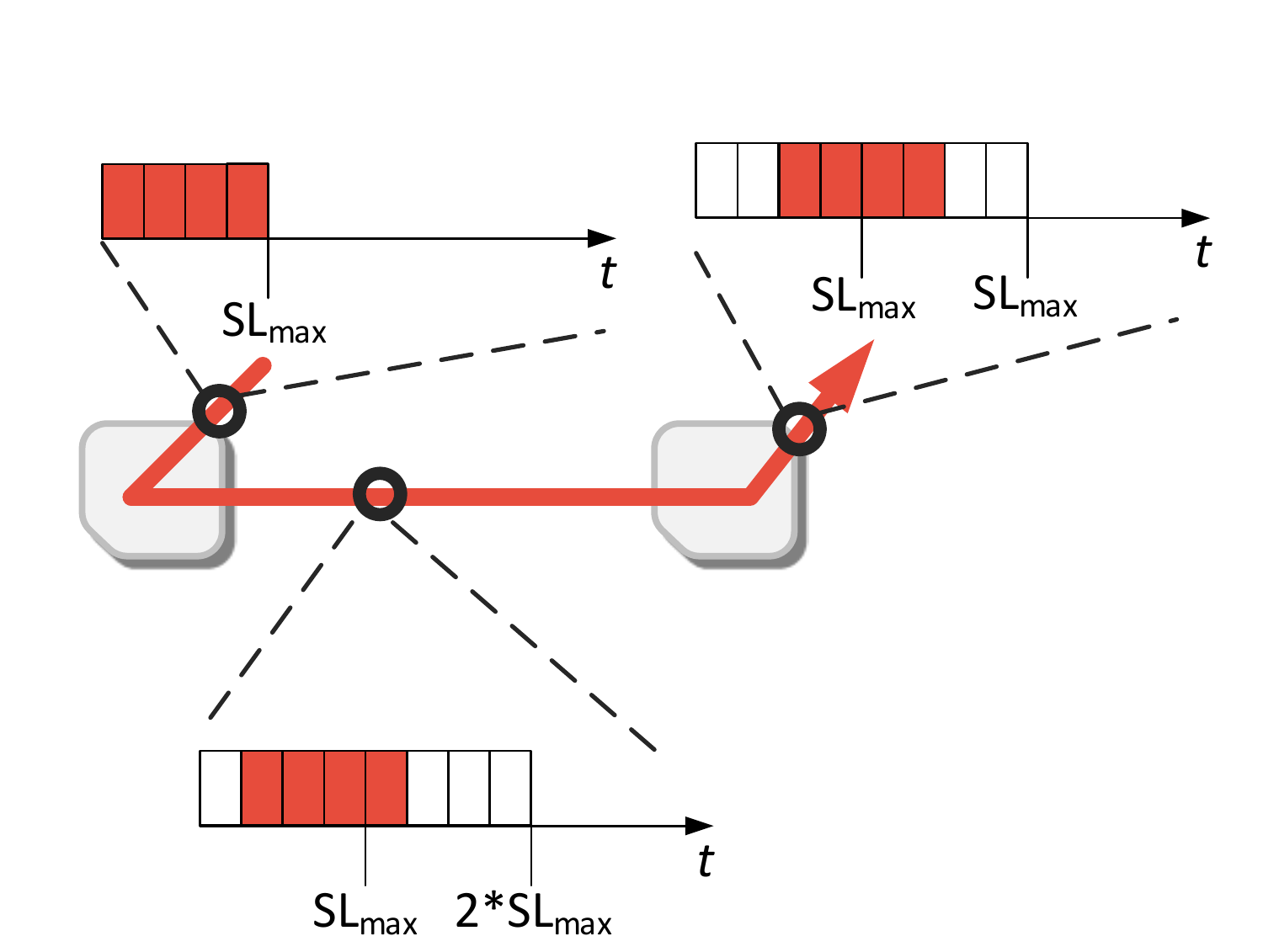}
\label{fig:wrr_best}
}
\subfigure[Worst Case]{
\includegraphics[width=0.45\columnwidth]{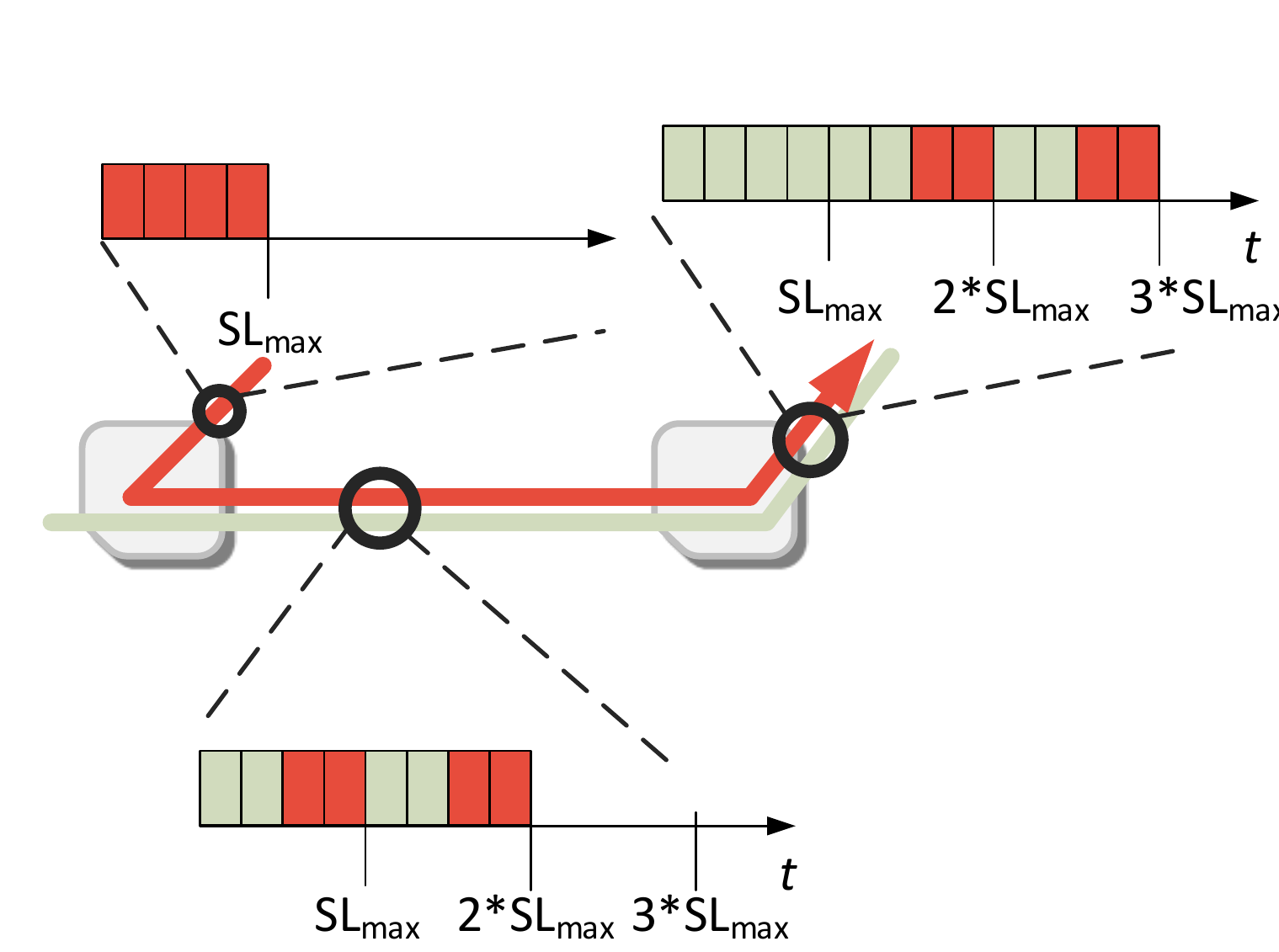}
\label{fig:wrr_worst}
}
\caption{Example for weighted round robin for the flows of flits of two messages $\mess_1$ and $\mess_2$. 
Periodically, $\SLmax=4$ time slots are available for transmission.
The red flow has $\SL(\mess_1)=2$ and consist of 4 flits ($\flits(\mess_1) = 4$).
In the best case (a) the latency only depends on $\flits(\mess_1)$, $\hop(\route(\mess))$, and $\rdelay$, i.e. Eq.~\eqref{eqn:nocdel:free}. 
Whereas, Eq.~\eqref{eqn:nocdel:blocked} describes the additional delay which can occur if only reserved time slots are available.
At each hop in $\left(\left\lceil\frac{\flits(\mess_1)}{\SL(\mess_1)} \right\rceil - 1\right)$ arbitration windows, the flits can be delayed by 
 \mbox{$(\SLmax-\SL(\mess_1)) \cdot \cycle$}.
Note that the position of the time slot can vary in each hop, while the number of time slots is assumed fixed per message.
Transmission can also use more time slots than actually reserved given there are unused time slots available. 
However, it is always guaranteed that at least the reserved time slots are available within the period.
}
  \label{fig:wrr}
\end{figure*}

Figure~\ref{fig:wrr} illustrates the best case and the worst case for communication latencies with examples. 
The best case corresponds to the case without any interference.
The message can utilize the whole scheduling interval $\SLmax$ and the transmission delay only depends on the message size $\flits$, the hop distance $hops$, and the router delay $\rdelay$ (see first summand in Eq.~\eqref{eqn:nocdel:free}).
The second summand in Eq.~\eqref{eqn:nocdel:blocked} gives the maximal delay possible by interference with other messages.
This interference can happen in $\left\lceil\frac{\flits(\mess)}{\SL(\mess)} \right\rceil - 1$ arbitration intervals and depends on the number of hops.

\subsection{Composable Task Scheduling}
\label{sec:scheduling}

Composability at \ac{PE} level is achieved   by temporally isolating the execution of tasks on it. 
Therefore, the processing time on a \ac{PE} is partitioned into \emph{service intervals} with fixed time duration. 
Within a service interval, tasks are scheduled exclusively.
We consider service intervals of equal length $\Snt$ on each \ac{PE} type $\type(\pe)$
\footnote{Generally, the service intervals on different \acp{PE} or \ac{PE} types could be of different lengths. 
Our approach could still work here, but for keeping notations simple, we make this assumptions.}.
\begin{revision2}
Transition between the scheduling of two tasks, i.e., task switching, takes place after each service interval $\Snt$. 
\end{revision2}
This incurs an \ac{OS} scheduling overhead after each interval denoted by $\SIos$. 
Service intervals are made available to the tasks in the \ac{PE}'s waiting queue in a round robin fashion. 
Each task is assigned with a fixed priority that determines the order in which intervals are allocated to tasks by the scheduler. 

\begin{figure}
 \centering
 \includegraphics[width=0.7\columnwidth]{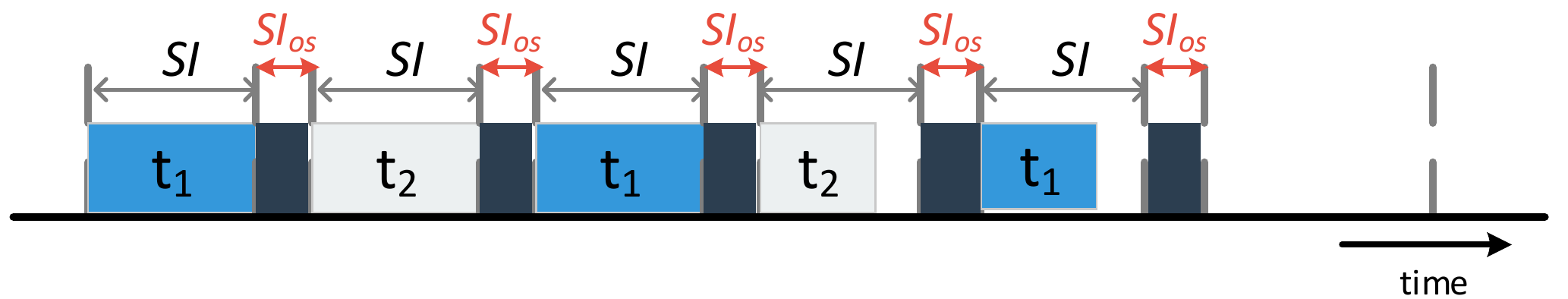}
 \caption{Scheduling of tasks (with temporal isolation) mapped on a PE. 
 	The tasks are executed in time slots with the length $\Snt$ and operating system overhead, e.g., task switching, is performed in $\SIos$.
 }
 \label{fig:pe_sched}
\end{figure}

This scheduling strategy is illustrated in Fig.~\ref{fig:pe_sched} for two tasks $t_1$ and $t_2$ in the ready queue of a \ac{PE}. 
The priority of task $t_1$ is higher than the priority of task $t_2$, i.\,e., $\prio(t_1) < \prio(t_2)$ (the lower the value the higher is the priority). 
So, task $t_1$ is assigned the first service interval. 
Allocation then proceeds by means of round robin scheduling. 
With the above scheduling mechanism, we develop a performance analysis method next to derive the worst-case execution latency of a task.

As initially stated, it is our goal to achieve a high utilization of the given many-core system despite having to isolate applications from each other in order to satisfy real-time constraints. 
The worst-case execution latency of a task basically consists of two parts: 
First, the \emph{worst-case execution time of the task without interference} $TL_{exec}(\task, \map(\task))$.
The proposed analysis  also considers an upper bound on the number of tasks that could share the same \ac{PE}, denoted by $K_{max}$.
Therefore, the second part is the \emph{worst-case interference} $TL_{inter}(\task, \map(\task))$ from other tasks that could possibly be mapped and scheduled on the same \ac{PE}.
Thus, the total worst-case execution latency ($TL(\task, \map(\task)$) of a task is given by
\begin{equation}
TL(\task, \map(\task)) = TL_{exec}(\task, \map(\task)) + TL_{inter}(\task, \map(\task))
\label{eqn:tl}
\end{equation}

As each task is executed in service intervals and considered to finish at the end of an interval, the value of $TL_{exec}(\task, \map(\task))$ is not necessarily equal to the \ac{WCET} $\wcet(\task, \map(\task))$. \footnote{\begin{revision2}
	If considering different frequencies the \ac{WCET} and the service intervals depend on the frequency $\frequency$, i.e., $\wcet(\task, \map(\task), \frequency)$, $\Snt(\frequency)$, and  $\SIos(\frequency)$.
	To simplify the formulae and w.l.o.g. we omit the frequency parameter in the remaining article. 
		\end{revision2} }
Rather, this value is given by
\begin{equation}
  TL_{exec}(\task, \map(\task))= \left\lceil \dfrac{\wcet(\task, \map(\task))}{\Snt} \right\rceil \times (\Snt+\SIos)
	.
  \label{eqn:twcet}
\end{equation}
\noindent The above expression is obtained from the fact that each task has to complete $\left\lceil {\wcet(\task, \map(\task))}/{\Snt} \right\rceil$ service intervals to finish its execution.
Moreover, each task execution incurs the \ac{OS} scheduling overhead $\SIos$ every time there is a switch into its service interval from the service interval of the previously scheduled task (cf. Fig.~\ref{fig:pe_sched}).

\begin{figure*}
	\centering
	 \subfigure[Maximal latency if predecessor is scheduled on the same \ac{PE}.]{
\label{fig:tlb1}\includegraphics[width=0.45\textwidth]{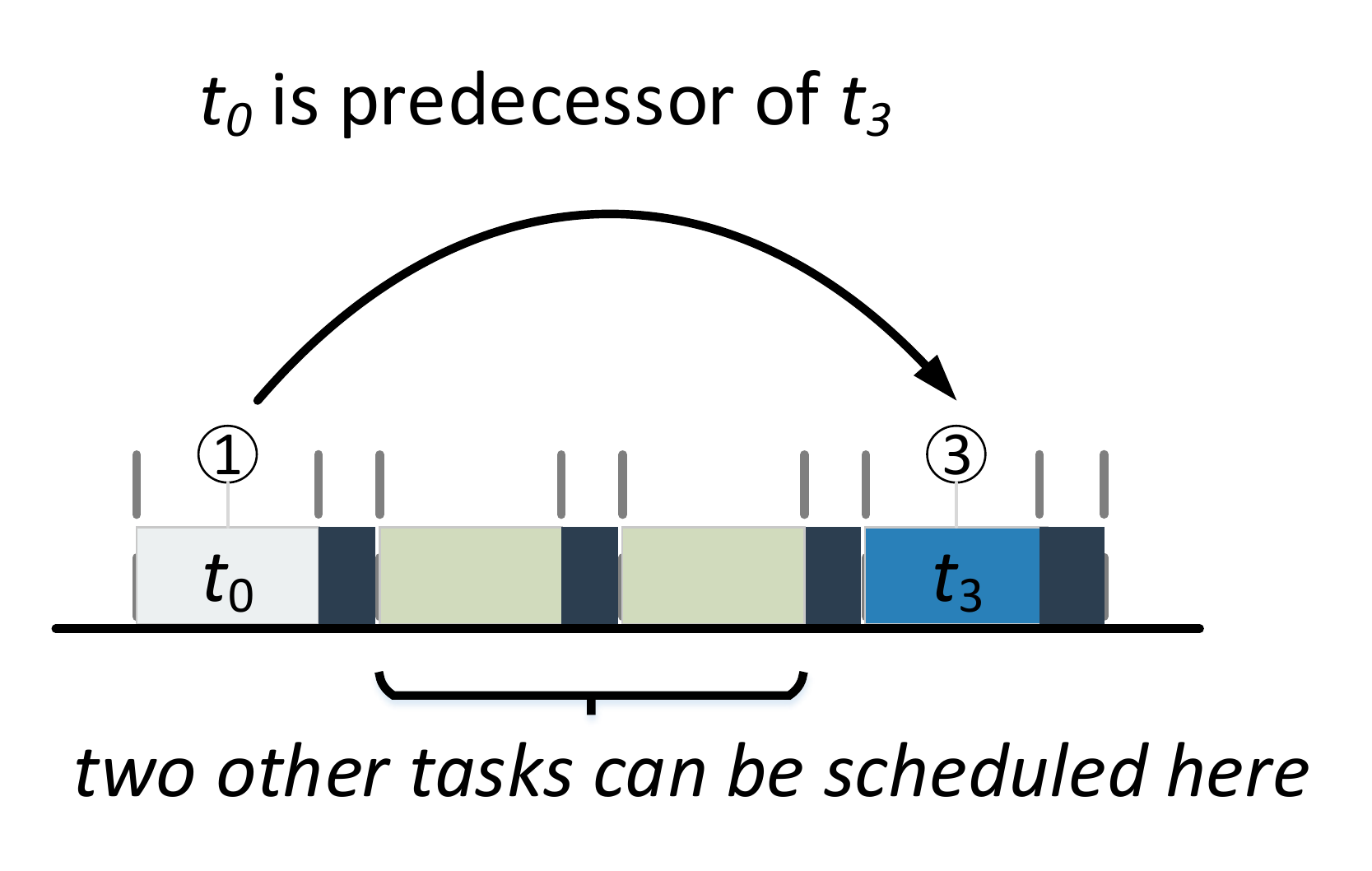}}
	 \subfigure[Maximal latency in all other cases where the introduced latency is solely dependent on $K_{max}$.]{
\label{fig:tlb2}\includegraphics[width=0.45\textwidth]{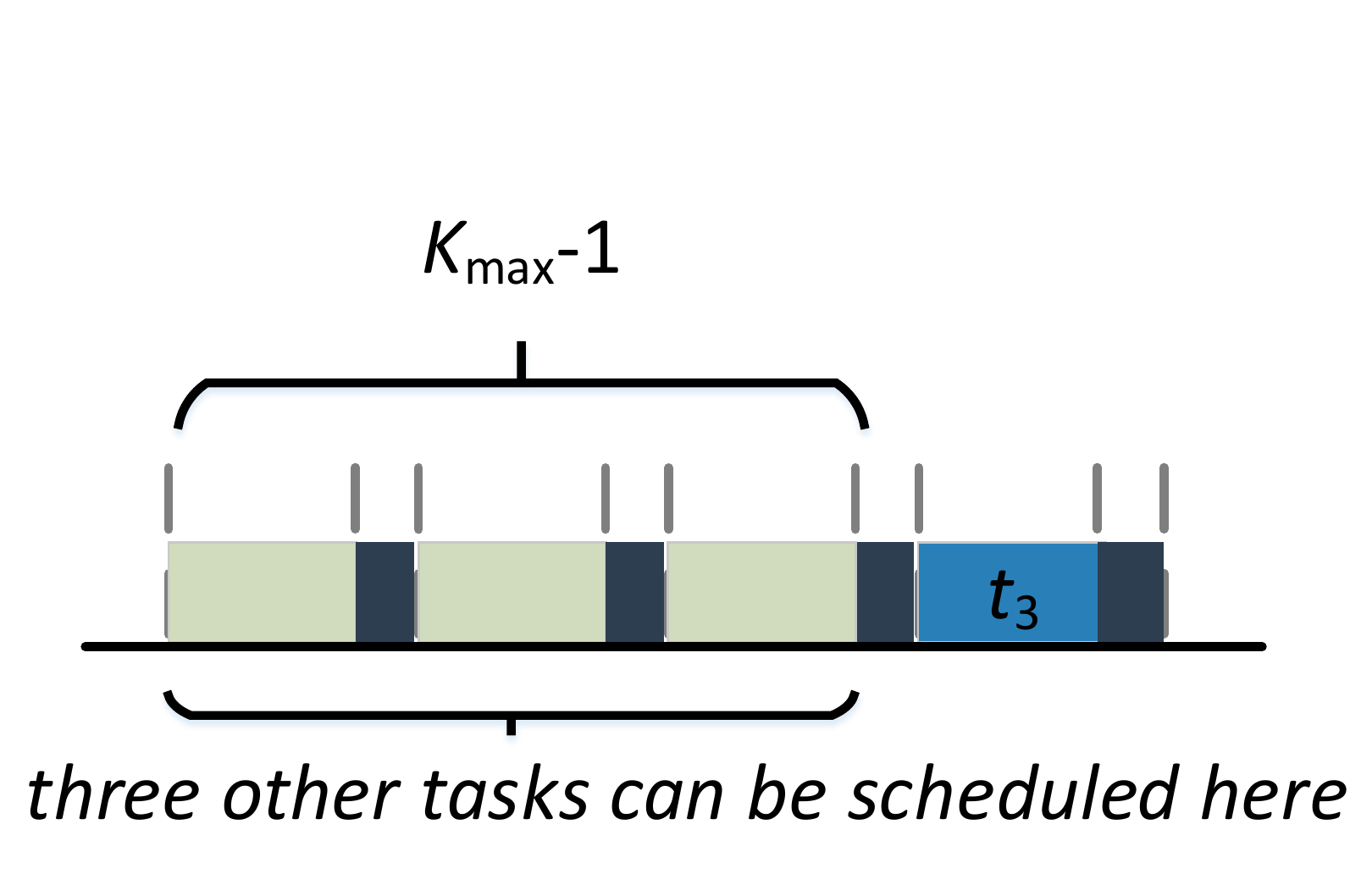}}
	\caption{Example of the two cases of Eq.~\eqref{eqn:tinterb}. The priorities of the tasks are annotated in circles.
	\label{fig:tlb_example}	}
\end{figure*}

The worst-case interference from other tasks consists of two components: 
the worst-case interference before $TL^{b}_{inter}(\task, \map(\task))$ and after $TL^{a}_{inter}(\task, \map(\task))$ the first service interval of $\task$, which is given by
\begin{equation}
TL_{inter}(\task, \map(\task)) = TL^{b}_{inter}(\task, \map(\task)) + TL^{a}_{inter}(\task, \map(\task))
.
\label{eqn:tinter}
\end{equation}

Recall $K_{max}$ being the maximum overall number of tasks allowed to be mapped onto a \ac{PE},
and let $pred(\task)$ be the predecessor of task $\task$ in the currently analyzed path of the task graph.
Then, worst-case interference before the first service interval is formulated as follows:
\begin{align}
    TL^{b}_{inter}(\task, \map(\task)) = 
    \begin{cases}
    \bigl(\prio(t)-\prio(pred(t))\bigr)\times \bigl(\Snt+\SIos\bigr), &\text{if $\map\bigl(pred(\task)\bigr)=\map(\task)$}\\
    \bigl(K_{max}-1\bigr)\times \bigl(\Snt+\SIos\bigr), &\text{otherwise}.    
  \end{cases}
  \label{eqn:tinterb}
\end{align}
If task $pred(\task)$ is mapped on the same PE as task $\task$, data is exchanged locally and the number of time intervals with length $(\Snt+\SIos)$ that $\task$ has to wait is $\prio(\task)-\prio(pred(\task))$, as exemplified in Fig.~\ref{fig:tlb1}.
On the other hand, if $pred(t)$ is mapped onto another PE, then the maximum interference is due to the service intervals of the possible number of other tasks ($K_{max}-1$) on the PE (see Fig.~\ref{fig:tlb2}).
This is because in the worst case, the message from $pred(t)$ would have to wait until the service intervals of all other tasks finish.

Worst-case interference after the first service interval is given by
\begin{equation}
  TL^{a}_{inter,t}(\task, \map(\task))= \left(\left\lceil \dfrac{\wcet(\task, \map(\task))}{\Snt} - 1 \right\rceil \right)\times tl_{inter}
  \label{eqn:tintera}
\end{equation}
where $tl_{inter} = (\Snt+\SIos)\times (K_{max}-1)$ is the maximal total interference from all the remaining possible tasks between two consecutive service intervals of task $\task$. 
The first part of the equation gives the number of service intervals of task $\task$ between which interference could happen (analogous to Eq.~\eqref{eqn:twcet}). 

The worst-case execution latency of task $t$ can then be calculated by inserting Eqs.~\eqref{eqn:twcet}--\eqref{eqn:tintera} into Eq.~\eqref{eqn:tl}.

\section{Design Space Exploration}
\label{sec:dse}
Due to our composability assumptions and using the performance analysis techniques presented in Section~\ref{sec:scheduling}, 
a \ac{DSE} for finding Pareto-optimal mappings is applied to each application individually. 
Here, multiple mapping candidates are generated per application with verified real-time properties and optimized objectives. 
The gain of this separation is that the  complexity of analyzing a single application is dramatically reduced over the exploration of a complete system with various application mixes. 

\begin{figure}[t]
	\centering
		\includegraphics[width=0.85\textwidth]{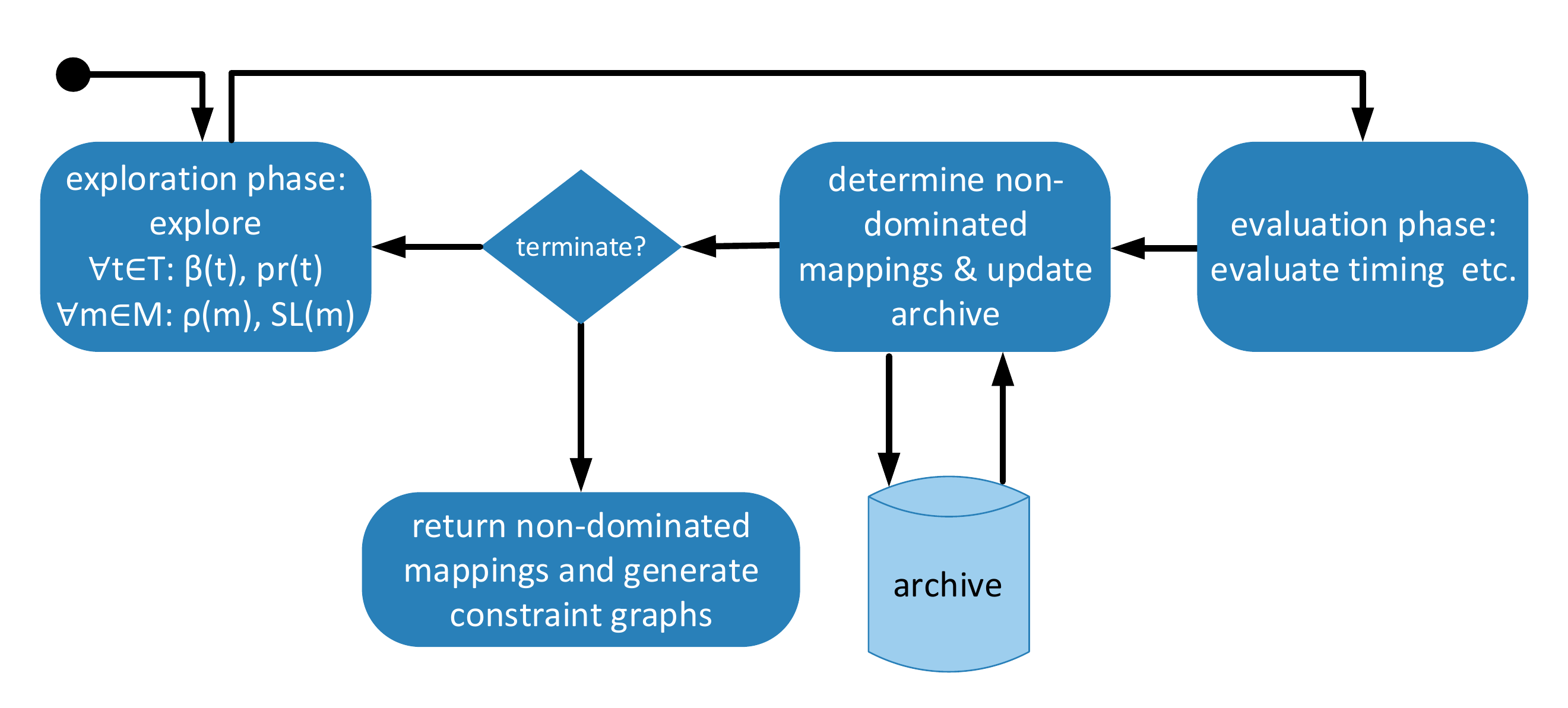}
	\caption{Flowchart of \ac{DSE} using \ac{EA}, including the iterative process of exploration and evaluation.}
	\label{fig:flowchart}
\end{figure}

To efficiently explore various mappings in our \ac{DSE}, we apply an approach that combines an~\ac{EA} with a Pseudo-Boolean solver~\cite{LukasiewyczGHT08}. 
The \ac{EA} constitutes an iterative optimization process: In the \emph{exploration phase} a set of new applications mappings is generated by applying genetic operators, and in the \emph{evaluation phase}, this set is evaluated by using analytical models (e.g. for timing the one presented in Section~\ref{sec:predictability}).
Both phases are iteratively carried out to obtain a set of better and better solutions over time. 
In each iteration, the best so far explored, non-dominated mappings are updated and stored in an archive and returned once the \ac{DSE} terminates (see Fig.~\ref{fig:flowchart}). 

%
Again, to enable the individual exploration of classes of optimal application mappings by means of a formal analysis, the concept of \emph{composability} is essential.
Composability ensures that the addition of a new application in the mix only has a bounded effect on the performance values obtained for each application that was analyzed completely in isolation without considering the execution behavior of any other application as this would fail due complexity reasons.

\subsection{Generation of Feasible Application Mappings}
We apply the composable scheduling techniques presented in the last section. 
This means that an application mapping during \ac{DSE} is generated by (a) determining a binding $\map(\task)$ of each task $\task \in \tasks$ and (b) determining a routing $\route(\mess)$ of each message $\mess\in\messages$. 
We consider deterministic \emph{xy-routing} for the messages in the \ac{NoC}. 
Routing of each message does, therefore, not have to be explored explicitly, as proposed in~\cite{GrafRGT14}, as it is implicit by the binding of the message's 
$\mess$ sending and receiving tasks.
In addition, also a priority $\prio(\task)$ has to be assigned to each task for scheduling tasks on the same \ac{PE}, and $\SL(\mess)$ has to be generated for the transmission of each message.

In our approach, unique priorities for each task mapped to the same \ac{PE} are assigned in the exploration phase.
In the evaluation phase, it is checked if the assignment is feasible.
Through a depth-first search, we identify if a task is a predecessor of another task on the same \ac{PE} and change the priorities if required.

Also, $\SL(\mess)$ has to be explored per message $\mess$.
To satisfy the minimal bandwidth requirements of the message,  $\SL(\mess)$ has to be at least $\frac{ \bandwidth(m)}{\capacity(l)} \cdot \SLmax$. 
By using a higher $\SL(\mess)$, however, the worst-case end-to-end latency $\EEL(\map, \route)$ may be reduced.
Therefore, the exploration interval of $\SL(\mess)$ is defined as follows:
\begin{equation*}
\SL(m) \in \Biggl[\left\lceil \frac{ \bandwidth(m)}{\capacity(l)} \cdot \SLmax \right\rceil, \SLmax \Biggr].
\end{equation*}

\begin{figure*}
	\centering
	 \subfigure[Infeasible mapping 1]{
\label{fig:mapping1}\includegraphics[width=0.3\textwidth]{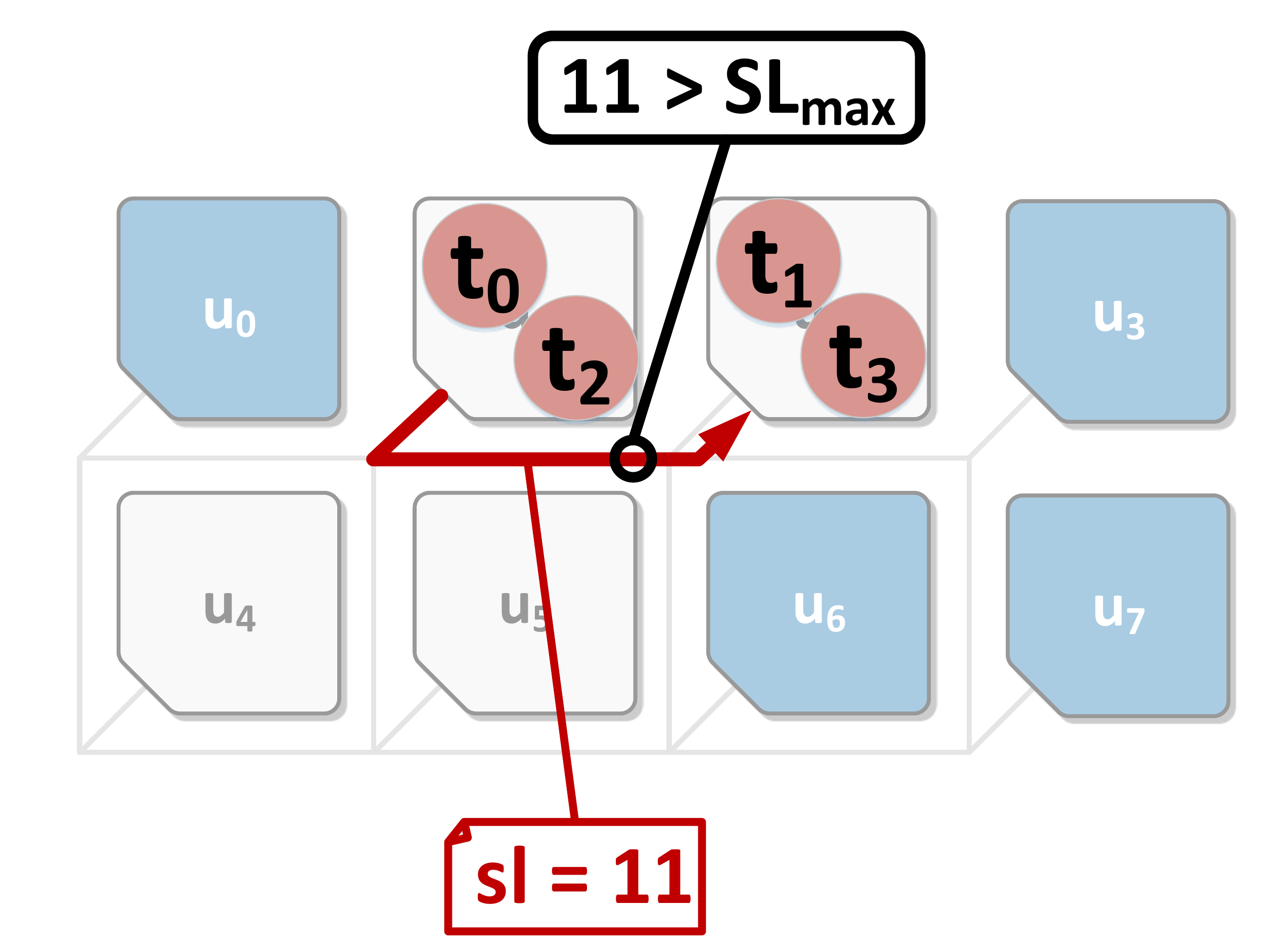}}
	 \subfigure[Infeasible mapping 2]{
\label{fig:mapping2}\includegraphics[width=0.3\textwidth]{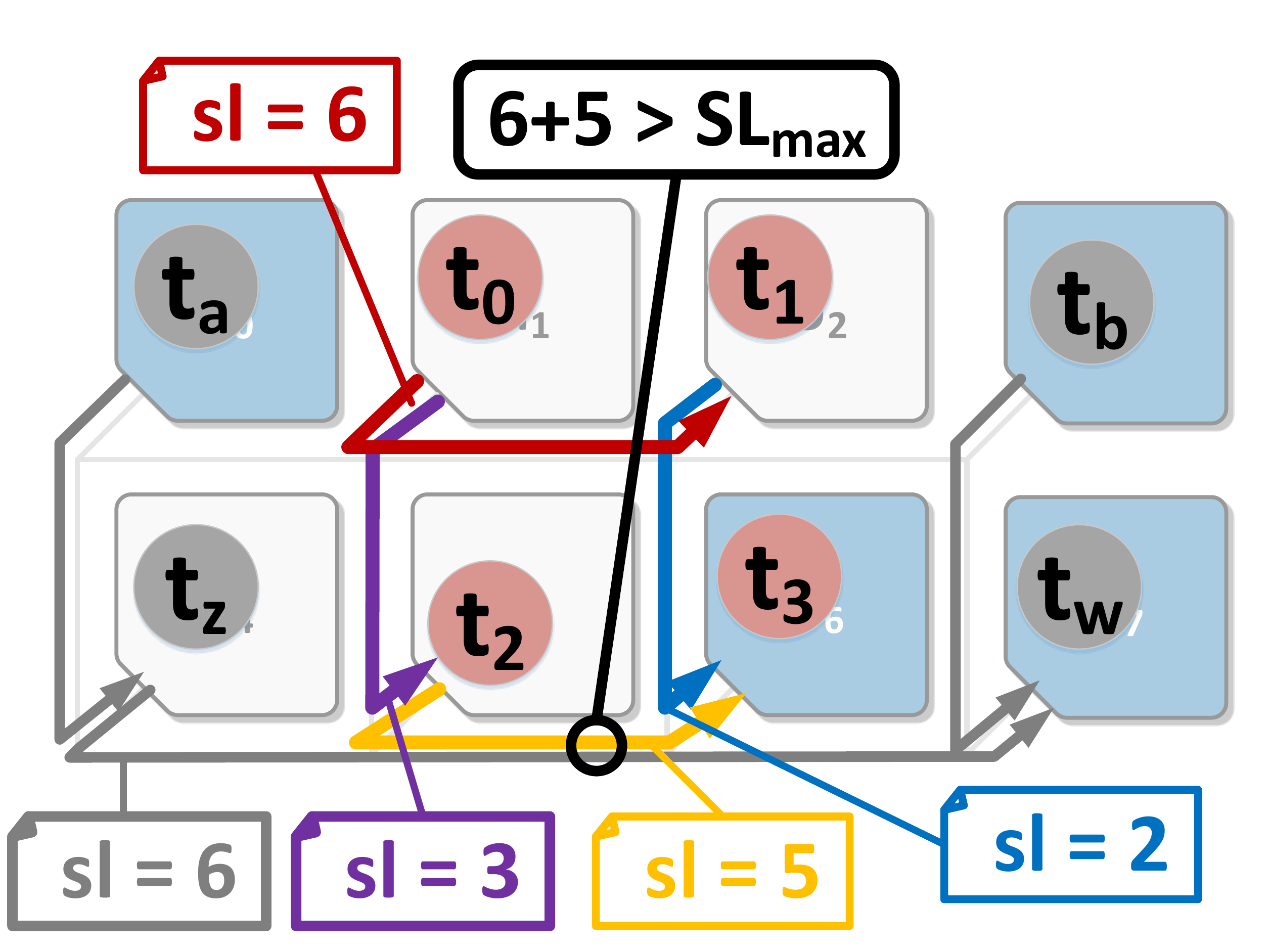}}
          \subfigure[Feasible mapping 3]{\label{fig:mapping3}
\includegraphics[width=0.3\textwidth]{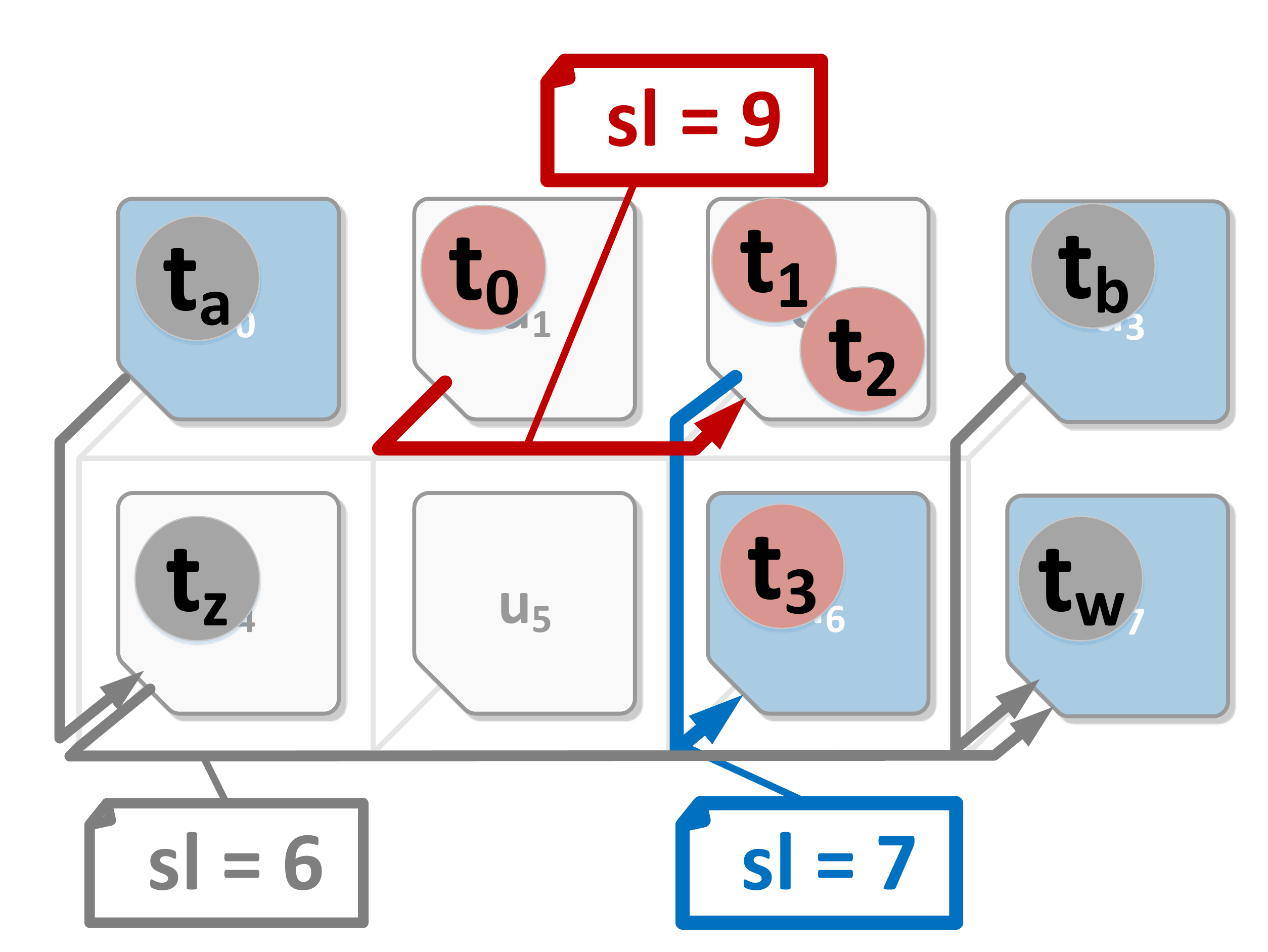} }	
	\caption{Several mappings of an application onto a \ac{NoC} with $\SLmax=10$. (a) and (b) are infeasible due to overutilization of shared 
resources, while (c) represents a feasible mapping \cite{WGWGT14}.}
	\label{fig:mapping}
\end{figure*}

Only \emph{feasible application mappings} are returned in the end. 
More formally, a mapping is feasible if the following conditions hold:
\begin{itemize}
\item 
    First, the worst-case end-to-end latency has to stay within the deadline:
    \begin{equation}
    \EEL(\map, \route) \le \deadline.
    \end{equation}
\item
    Second, no \ac{PE} is overutilized. 
    Meaning that the load induced by all tasks mapped onto the same \ac{PE} stays below $100\,\%$: 
    \begin{align}
    \frac{\sum_{\overset{\task \in \tasks:}{\map(\task)=\pe}} \left\lceil \frac{\wcet(\task, \map(\task))}{\Snt(\task)} \right\rceil \cdot 
    (\Snt(\task)+\SIos)}{ \period} \leq 1 ~,~~~\forall \pe \in \pes
    .
    \end{align}
\item
    Finally, no communication link is overutilized. 
    This means that  $\SL(\mess)$ of all messages that are sent over the same route (same source \ac{PE} and target \ac{PE}) do not exceed the overall available budget of time slots $\SLmax$. 
    Let $\messages_\route = \{ \mess ~|~ \mess, \mess^\prime \in \messages: \route(\mess) = \route(\mess^\prime) \}$ be the set of messages that are sent over the same route. 
    This constraint is then formulated as follows:
      \begin{align}
      \sum_{\mess \in \messages_\route}     
        \SL(\mess)  \le \SLmax,~~ \forall \messages_\route
      \end{align}    
      An example of such an infeasible mapping due to overutilization of a link is illustrated in Fig.~\ref{fig:mapping1}.
\end{itemize}

\par

\subsection{Optimization Objectives and Evaluation}\label{sec:objectives}

Our \ac{DSE}   considers multiple objectives related to non-functional properties. 
As modern embedded system have strict energy budgets, it is essential to minimize the energy consumption of application mappings. 
Therefore, we include energy consumption minimization as one objective in the \ac{DSE} (Objective I). 
This \emph{maximal energy consumption} $E_{OV}$ that is going to be minimized may be the sum of the energy consumed by the \acp{PE} $E_{PE}$ and energy which is used to route the message over the \ac{NoC} $E_{NoC}$:		
\begin{align}
E_{OV}&=E_{PE}+E_{NoC} \\
E_{PE}&= \sum_{t \in T} \bigg (  \power  \Big( \type \big ( \map(t) \big ) \Big ) \cdot \wcet \big (t, \map(t) \big ) \bigg)
\label{eq:energy}
\end{align}
The maximal energy consumed in the \ac{PE} is the product of the \ac{WCET} of the task on the mapped \ac{PE} and the maximal power consumption  $\power(\res)$ for the given resource type which is derived by the function $\type(\pe)$.
The energy consumed by the communication infrastructure for a message $\mess$  is directly proportional to the number of hops and used links.
We derive $E_{NoC}$ from the \ac{NoC} energy model in~\cite{Hu:2003,Wolkotte:2005}:

\begin{align}
E_{NoCbit}^{ \route( \mess ) } =&  \hop \big (  \route  ( \mess ) \big )  \cdot E_{Sbit} +  \Big ( \hop \big ( \route(\mess) \big )-1 \Big ) \cdot E_{Lbit} \label{eq:nocenergy1}\\
E_{NoC} =& \sum_{  \mess \in \messages} \big ( E_{NoCbit}^{\route(\mess)} \cdot \size(\mess) \big).
\label{eq:nocenergy}
\end{align}
In Eq.~\ref{eq:nocenergy1}, $E_{Sbit}$ is the energy consumed per bit inside the router, $E_{Lbit}$ is the energy consumed on a link, and $\size(\mess)$ is the size of the message in bits.

Contrary to conventional exploration, the outcome of the \ac{DSE} will not be used to encode a concrete task and communication assignment to be selected by the \ac{RM} but rather a \emph{class of mappings}.
More details are elaborated in Section~\ref{sec:runtime}. 
In order to find mappings  that allow a greater  run-time flexibility, we therefore also include objectives that quantify the resource overhead and flexibility of an application mapping as follows:

The overall number of messages routed over the \ac{NoC} should be minimized (Objective II).
The reason is that, if two communicating tasks are mapped to the same \ac{PE}, they can exchange their data through local memory and hence $\route = \emptyset$.
This does not burden the \ac{NoC} infrastructure. 
Consequently, congestion on \ac{NoC} links is reduced, making it more likely to map this operating point at run time.

Another two objectives are the maximization of the average and the minimal hop distances (Objective III and IV).
Again, here the idea is to increase flexibility by giving preference to routings that are more likely to be feasibly routed during run time: the longer routes are allowed to be, the less mapping restrictions exist.

As the targeted architecture is heterogeneous, different \ac{PE} types may be selected for the execution of the tasks.
Only minimizing the overall number of allocated \acp{PE} without differentiating between their resource types,  will result in the generation of suboptimal mappings, e.g., by always using the same \ac{PE} type such as a powerful core which can execute many tasks within the application's period. 
However, if now during run time all instances of this \ac{PE} type are occupied, no more operating points could be embedded in the system.
To thwart this, we minimize the number of allocated \acp{PE} \emph{per resource type} to generate diverse operating points (Objective V).

Our \ac{DSE} therefore performs a multi-objective optimization, with an overall of five objectives.
This results not in a single optimal, but in multiple Pareto-optimal application mappings that trade-off between the different objectives.
Such a Pareto front is illustrated in Fig.~\ref{fig:cluster_graphs} for two objectives.

\section{Run-time Constraint Solving}
\label{sec:runtime}
\begin{figure}[t]
 \centering
 \includegraphics[width=0.90\columnwidth]{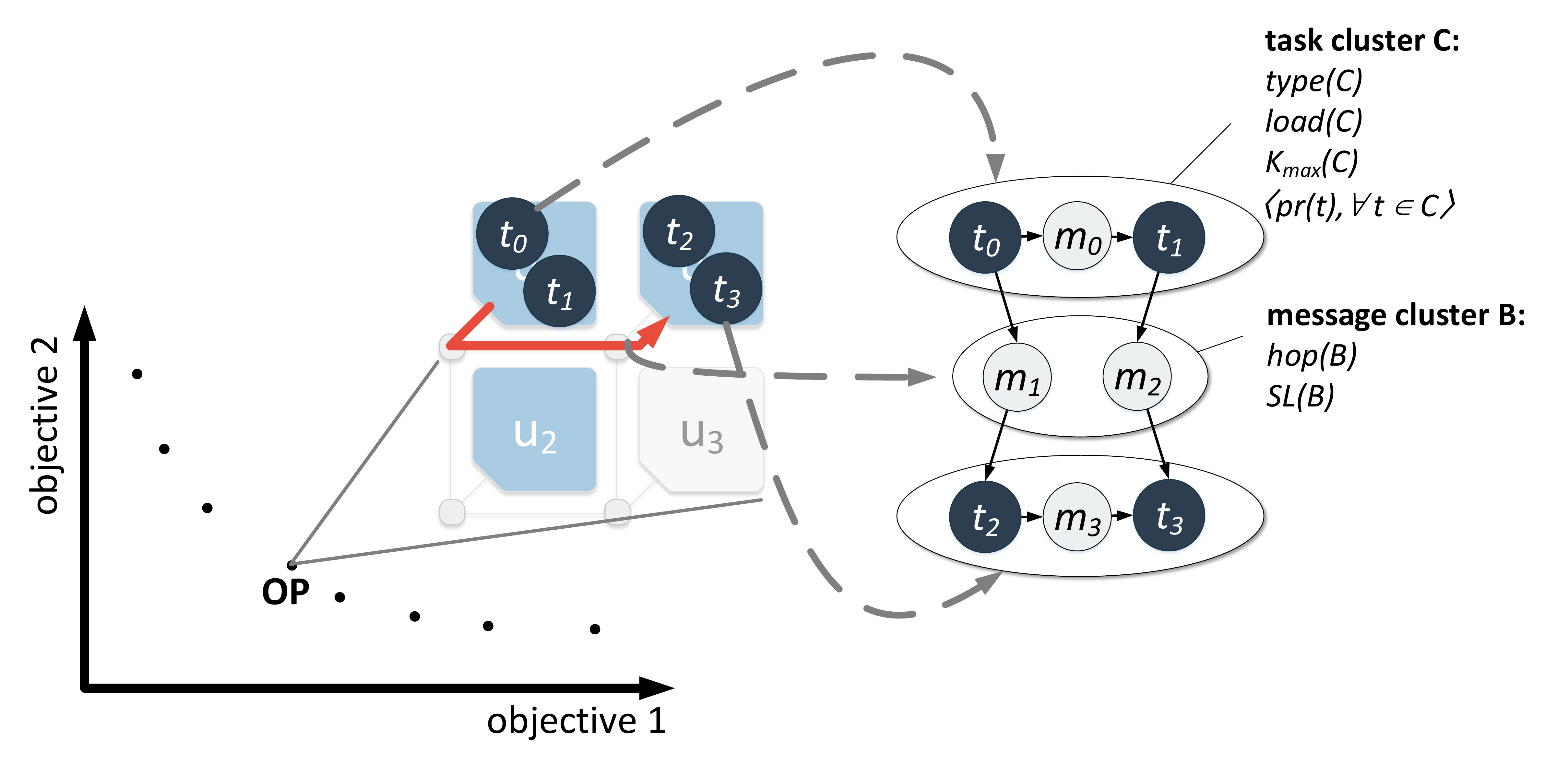}
  \caption{Representation of  the Pareto-front of explored mappings (left) and the concept of a constraint graph (right) of an explored mapping containing two task clusters and one message cluster  with annotated constraint information.\label{fig:cluster_graphs}}
\end{figure}

The Pareto-optimal mappings generated by the \ac{DSE} are handed over to the \ac{RM}. 
Yet each \ac{DSE} mapping corresponds to a fixed assignment of tasks to concrete resources in the architecture.
However, in architectures with a multitude of equal resources, numerous equivalent mappings may exist. 
Therefore, we transform the application mapping (provided by $\map$ and $\route$) into a \emph{constraint graph} $\Gconstraint$ as exemplified in Fig.~\ref{fig:cluster_graphs} right. 
This graph represents a full class of symmetrical feasible mappings within the \ac{NoC} which are all equivalent to the application mapping that was actually determined and analyzed during \ac{DSE}. 
Consequently, all analyzed properties---particularly real-time properties---also apply for these symmetrical mappings.

\subsection{Constraint Graphs}

As illustrated in Fig.~\ref{fig:cluster_graphs}, the vertices $V_C= \taskclusters \cup  \commchannels$ of a constraint graph are composed of task clusters belonging to the set $\taskclusters$ and message clusters belonging to the set $\commchannels$. 

Each \textit{task cluster} $\taskcluster \in \taskclusters$ represents a set of tasks that are mapped to the same \ac{PE}, so that $\forall \task, \task^\prime \in \taskcluster: ~\mapDSE(\task)=\mapDSE(\task^\prime)$. 

Each task cluster is annotated with $\type_{CG}(\taskcluster) \in \RES$, specifying the \ac{PE} type onto which the tasks are mapped, 
and furthermore, with load $load(\taskcluster)$ induced by the tasks on this \ac{PE}:
\begin{equation}
load(\taskcluster)=\sum_{\forall \task \in \taskcluster } 
 \left\lceil \dfrac{\wcet(t,\map(t))}{\Snt} \right\rceil \times \dfrac{(\Snt+\SIos)}{\period}
\end{equation}
Also, the scheduling information is annotated to the task cluster, i.e., the maximum number $K_{max}(\taskcluster)$ of tasks allowed on the \ac{PE} for scheduling and the priorities $\prios$ of all its tasks. 

Each \textsl{message cluster} $\commchannel \in \commchannels$ represents a set of all messages which are routed along the same path in the \ac{NoC} between two such task clusters, so that $\forall \mess, \mess^\prime \in \commchannel: ~\route(\mess)=\route(\mess^\prime)$. 
Each message cluster is annotated also with the routing information, i.e. the accumulated $\SL(\commchannel) = \sum_{\mess \in \commchannel} SL(\mess)$ and the hop distance $hop(\commchannel) = \hop(\route(\mess))$ between the sending and the receiving task clusters of messages $\mess \in \commchannel$. 

{
\subsection{Serializing Operating Points}
To hand over the set of operating points to the \ac{RM} the data has to be serialized.
This includes the constraint graph as well as the values for the explored objectives.
The memory requirement for these tuples can be calculated as follows:
\begin{equation}
size_{OP}= size _{CG} \cdot n_{obj}\cdot size_{obj}
\end{equation}
Where $size _{CG}$ is the memory requirement of the serialized constraint graph, $n_{obj}$ is the number optimized objectives and $size_{obj}$ is the memory requirement of one objective value.
For serializing the constraint graph the graph needs to be traversed and all task clusters, all message clusters, and all edges have to be serialized.
\begin{equation}
size _{CG} = |\taskclusters| \cdot size_{\taskcluster} +  |\commchannels| \cdot size_{\commchannel} + |E_C| \cdot size_{e}
\end{equation}
For a task cluster, a constraint graph unique ID, the number of tasks $|\taskcluster|$ , the load $load(\taskcluster)$,  the number of additional tasks $K_{max}$, and the priorities of the tasks need to be saved.
The serialized message clusters include an ID, the hop constraint $hop(\commchannel)$, and the SL constraint $\SL(\commchannel)$.
With  $n_{obj}=7$, $size_{obj}=\SI{4}{\byte}$, $ |\taskclusters|=10$, $size_{\taskcluster}=\SI{10}{\byte}$, $ |\commchannels|=8$, $size_{\commchannel} =\SI{6}{\byte}$,  and $|E_C|=12 $ $size_{e}=\SI{4}{\byte}$ would result in $size _{CG}=\SI{196}{\byte}$ and $size_{OP}=\SI{224}{\byte}$.
}


\subsection{Run-time Mapping of Constraint Graphs}

\begin{figure}[t]
	\centering
		\includegraphics[width=0.7\columnwidth]{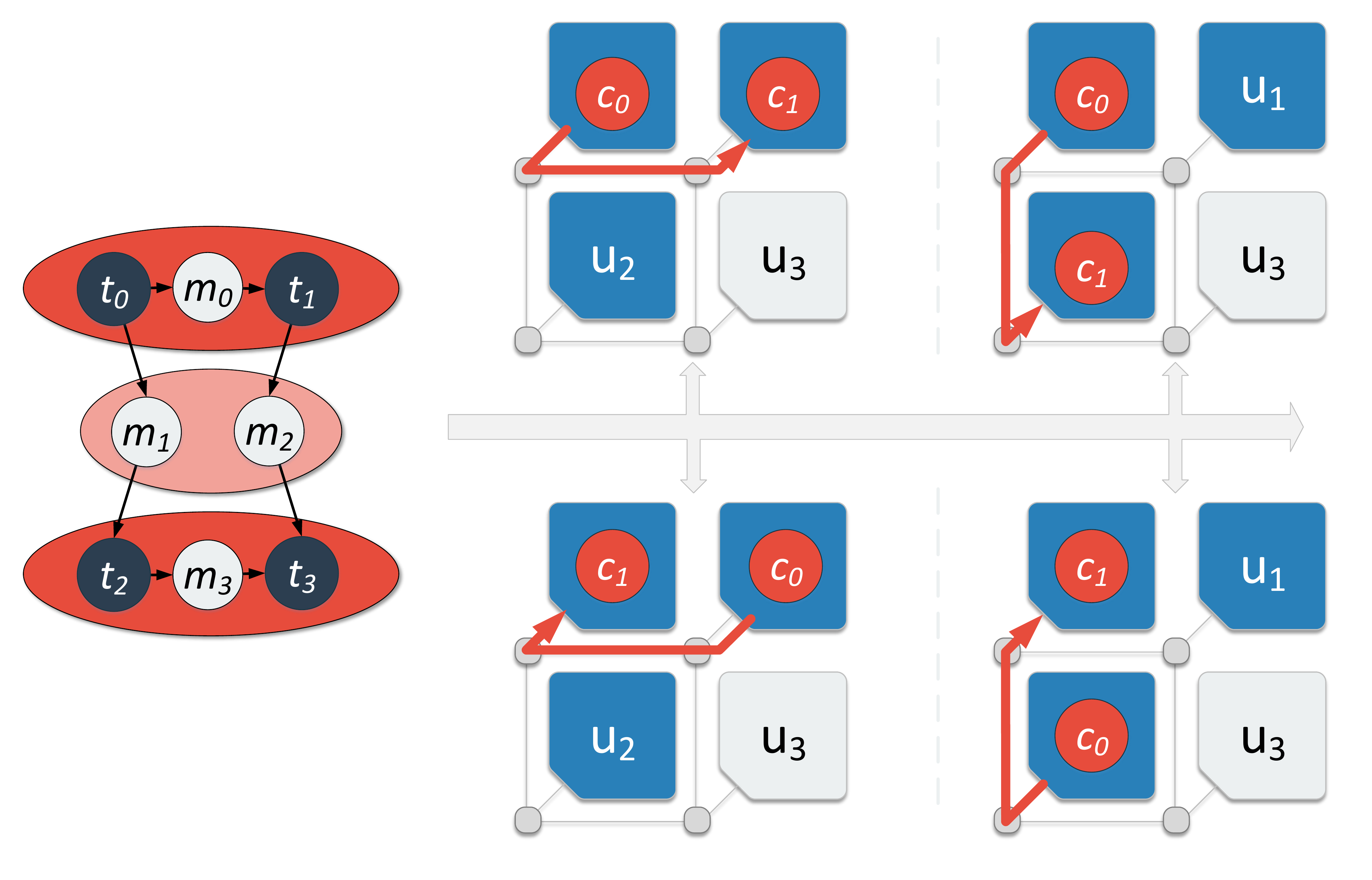}
	\caption{\label{fig:symmetric mappings}One constraint graph represents multiple feasible mappings with the same characteristics and bounds on end-to-end latency and energy consumption.}
\end{figure}

{The main task of the \ac{RM} is to select  a suitable operating point of the application that should be executed and do the actual  run-time application mapping.
In principal, the \ac{RM} can select any  operating point out of all found points which fulfills the application's requirement, e.g. performance.\footnote{{A methodology to order and select \acp{OP} for energy-efficient mappings can be found in \cite{WildermannWT15}.}}
To fulfill system requirements, e.g. utilization, the \ac{RM} may also re-map an already mapped application to another operating point.
The step of \emph{run-time} application mapping itself is to find a concrete application mapping based on the notation of a constraint graph $\Gconstraint$ and the architecture $\Garch$ 
by (a) \emph{binding} each task cluster to a \ac{PE}, i.e. $\mapCG{:}~\taskclusters~{\rightarrow}~\pes$, and (b) \emph{routing} each message cluster over a route of consecutive links, i.e., $\routeCG: \commchannels \rightarrow 2^{\links}$.}\footnote{Note the difference of the binding and routing to $\map$ and $\route$ during \ac{DSE}.} 
Instead of mapping the task graph $\Gapp$ onto the architecture, mapping the constraint graph $\Gconstraint$ has a lot of advantages:
As tasks are clustered to a task cluster and  messages to a message cluster, it is evident that $|\taskclusters| \leq |\tasks|$ and $|\commchannels| \leq |\messages|$.
In consequence, the size of the graph that needs to be mapped during run time is smaller than the original size of the task graph.
Second, the constraint graph also is a very compact representation of possibly multiple \emph{symmetrical} run-time mappings.
This basic idea is illustrated in Fig.~\ref{fig:symmetric mappings}, where one constraint graph can be feasibly mapped in multiple ways while guaranteeing the analyzed quality bounds. 
Third, time-consuming analysis is performed at design time.
The analyzed properties apply for a mapped constraint graph due to the composability of our approach.

A feasible mapping of a constraint graph has to satisfy the following constraints:
 First, the \emph{routings} of all message clusters $\commchannel \in \commchannels$ have to fulfill constraints \ref{const:hop} and~\ref{const:sl}:
\begin{enumerate}[label=\emph{C.\arabic*}]
   \item Routing $\routeCG(\commchannel)$ has to provide a connected route of links between $\mapCG(\taskcluster_1)$ and $\mapCG(\taskcluster_2)$, 
i.e., the target \acp{PE} of its sending and receiving task clusters are $\mapCG(\taskcluster_1)$ and $\mapCG(\taskcluster_2)$, respectively, 
with $(\taskcluster_1,\commchannel), (\commchannel,\taskcluster_2) \in E_C$.  \label{const:hop}
   The hop count of this route must not exceed the given maximal hop count associated with the message cluster:
    \begin{align}
	\hop\left( \routeCG(\commchannel) \right) \le hop(\commchannel)     
   \end{align}
   \label{const:hops}
   \item Let $\mathcal{\commchannels}$ denote the set of all already routed message clusters in the system.    \label{const:sl}
   The accumulated $\SL(\commchannel)$  of the messages routed over each link $l \in \routeCG(\commchannel)$ must not exceed the maximal number of time slots $\SLmax$:
\begin{align}
\SL(\commchannel) +  \smashoperator[r]{\sum_{\overset{\commchannel^\prime \in \mathcal{\commchannels}:}{l \in 
\routeCG(\commchannel^\prime)}}} \SL(\commchannel^\prime)  \le \SLmax,~~ \forall l \in \routeCG(\commchannel)
\end{align}
{
Figure~\ref{fig:mapping2} gives an example where this constraint is violated resulting in an infeasible run-time mapping.}
\end{enumerate}

\noindent Second, the \emph{bindings} of all task clusters $\taskcluster \in \taskclusters$ have to fulfill constraints \ref{const:type}--\ref{const:k}:
\begin{enumerate}[label=\emph{C.\arabic*}]
  \setcounter{enumi}{2}
 \item The resource type of the target \ac{PE} has to be the same as is required for the task cluster: \label{const:type}
\begin{align}
    \type\bigl(\mapCG(\taskcluster)\bigr)=\type_{CG}(\taskcluster)
\end{align}
 \item Let $\mathcal{\taskclusters}$ denote the set of task clusters that are already bound. 
 The load induced by all task clusters which are mapped on a target \ac{PE} $\mapCG(\taskcluster)$ together with the load of the new task cluster $\taskcluster$ must not exceed $100\%$: \label{const:load}
 \begin{align}
     load(\taskcluster) +  \smashoperator[r]{\sum_{\overset{\taskcluster^\prime \in \mathcal{\taskclusters}:}{\mapCG(\taskcluster^\prime) =%
\mapCG(\taskcluster)}}} load(\taskcluster^\prime) \le 1
 \end{align}
 \item The overall number of tasks bound on a  target \ac{PE} must not exceed the maximal numbers $K_{max}$ allowed for feasibly 
scheduling any task cluster on the \ac{PE} according to its performance analysis results:
\label{const:k}
\end{enumerate}
\begin{align}
|\taskcluster| + \smashoperator[r]{\sum_{{\overset{\taskcluster^\prime \in \mathcal{\taskclusters}:}{\mapCG(\taskcluster^\prime) = 
\mapCG(\taskcluster)}}}} |\taskcluster^\prime|
\le
\smashoperator[r]{\min_{\overset{\taskcluster^\prime \in \mathcal{\taskclusters}:}{\mapCG(\taskcluster^\prime) = \mapCG(\taskcluster)}}} 
\left\{
  K_{max}(\taskcluster),
  K_{max}(\taskcluster^\prime)
\right\}
\end{align}

\begin{figure}[t]
	\centering
		\includegraphics[width=0.60\columnwidth]{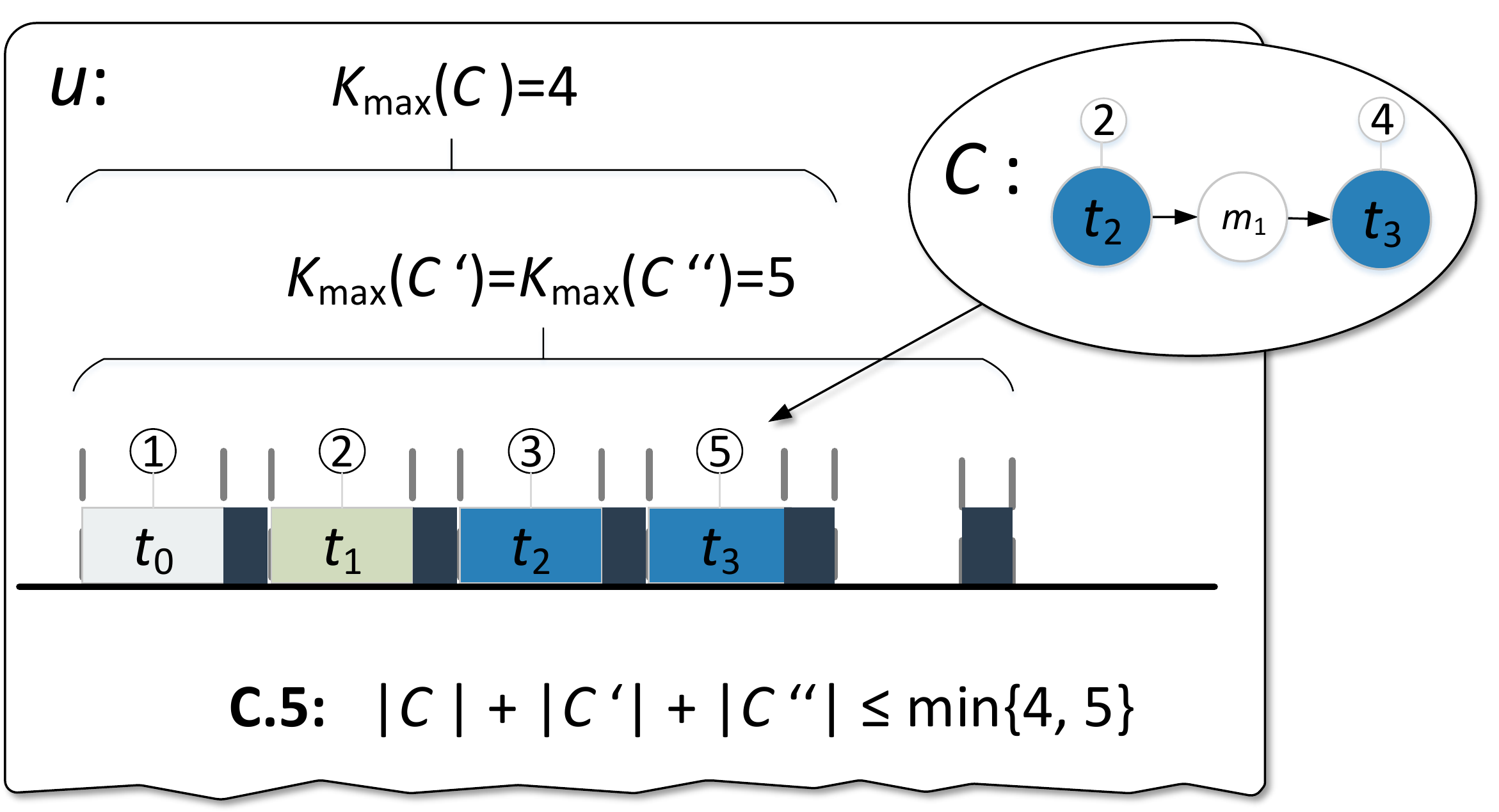}
	\caption{Example of a binding of a task cluster $\taskcluster = \{\task_2, \task_3\}$ to $\pe$. 
	The maximal number of tasks allowed on a \ac{PE} for scheduling $\taskcluster$ is $K_{max}(\taskcluster)=4$. 
	The tasks from task clusters $\taskcluster^\prime =\{t_0\}$ and $\taskcluster^{\prime\prime} =\{t_1\}$, $\taskcluster^{\prime}, \taskcluster^{\prime\prime} \in \mathcal{\taskclusters}$ are already present at $\pe$ and support a maximum task number of $K_{max}(\taskcluster^\prime)=K_{max}(\taskcluster^{\prime\prime})=5$. 
	After mapping $\taskcluster$, no further tasks can be mapped onto $\pe$ due to constraint~\ref{const:k}.
	The priorities (annotated in circles) of the tasks in $\taskcluster$ are updated to \textcircled{3} and \textcircled{5} in order to keep the priorities on $\pe$ unique.}
	\label{fig:new_mapping}
\end{figure}

In case of a spatial isolation, only constraint~\ref{const:type} and the absence of other tasks on $\mapCG(\taskcluster)$ 
would be sufficient to guarantee the worst-case latency (see Eq.~\eqref{eqn:tl}) as only tasks of one task cluster would be mapped together onto the same \ac{PE}. 
However, when applying temporal isolation, all constraints need to hold. 
{
Figure~\ref{fig:mapping3} exemplifies a feasible run-time mapping which fulfills all mentioned constraints.}
If all constraints are fulfilled but the priority ranges of the tasks in $\taskcluster$ and in $\taskcluster^\prime$ overlap, the priorities of $\taskcluster$ are shifted after mapping to keep them unique on the \ac{PE}. 
An example of this priority assignment and constraint~\ref{const:k} can be found in Fig.~\ref{fig:new_mapping}.

\subsection{Backtracking Algorithm}

To find a mapping which satisfies all the five constraints given a constraint graph and to solve the corresponding \acl{CSP}\footnote{The mapping of the constraint graph is a variant of task mapping which can be modeled as a bin-packing problem which is NP-complete \cite{bbw08}.
}, we propose a \emph{backtracking algorithm} as shown in Algo.~\ref{algo:backtracking} and is an extension of the algorithm presented in~\cite{WGWGT14}.
{
This algorithm is executed for each application that should be started on the system.
}
This algorithm starts with $\mathcal{A}=\emptyset$ and then searches recursively for a valid variable assignment for $\mathcal{A}$. 
As the backtracking algorithm would search exhaustively through all possible variable assignments,  a timeout can be chosen to determine the maximal run time of the algorithm.
This  condition is checked in line~\ref{algo:timeout}, and returns an empty set if the maximal time has elapsed since the initial start of the backtracking algorithm for one operating point.
In line~\ref{algo:BT:select}, the next task cluster to map is selected, and in line~\ref{algo:BT:domain} the domain $D_\taskcluster$ containing all target \acp{PE}
which fulfill \ref{const:hops} and \ref{const:type} is created.
In lines~\ref{algo:BT:testStart}~to~\ref{algo:BT:testEnd}, the remaining constraints are checked when trying to map $\taskcluster$ to the selected \ac{PE} $\pe$. 
We use xy-routing to obtain routes $L_\commchannel$ deterministically for all message clusters sent or received by $\taskcluster$ and which communication partners are already mapped.

 \begin{algorithm}[ht]
 \label{algo:backtracking}
     {\tt backtrack}($\mathcal{A} $, $\Gconstraints$, $\Garchs$)\\
    \nl \If{($\mathcal{A} $ is complete)\label{algo:BT:cond1}}{
       \textbf{return} $\mathcal{A} $\;
     }
		\If{(timeOut)\label{algo:timeout}}{
       \textbf{return $\emptyset$} \;
     }
     $\taskcluster$ = getNextTaskCluster($T_C$)\;\label{algo:BT:select}
     $D_\taskcluster$ = find \acp{PE} satisfying Constraints \ref{const:hops} and \ref{const:type}\;\label{algo:BT:domain}
     \For{\textbf{each} $\left( \pe \in D_\taskcluster \right)$\label{algo:BT:testStart}} {
		$\mapCG(\taskcluster)$ {and} $\route(\pred(\taskcluster))$\;
 	\If{ $\langle \mapCG(\taskcluster), \routeCG \rangle$ fulfills Constraints \ref{const:sl}, \ref{const:load},\ref{const:k}}{
 	    $\mathcal{A} ^\prime$ = {\tt backtrack}($\mathcal{A} \cup \langle \taskcluster, \mapCG(\taskcluster), \routeCG \rangle$, $\Gconstraints$,
$\Garchs$)\;\label{algo:BT:recursion}
 	    \If{($\mathcal{A} ^\prime \ne \emptyset$)}{
 	      \textbf{return} $\mathcal{A} ^\prime$\;
 	    }\label{algo:BT:testEnd}
 	}
     }
      \textbf{return} $\emptyset$\;
 \caption{Backtracking algorithm for finding a feasible constraint graph mapping.}
 \end{algorithm}

\section{Experiments}
\label{sec:experiments}

We use task graphs from the \emph{Embedded System Synthesis Benchmarks Suite (E3S)}~\cite{e3s} for our experiments.
These applications stem from various embedded domains like automotive ($18$ tasks), telecommunication ($14$ tasks), consumer ($11$ tasks), 
and networking ($7$ tasks). 
The values for energy consumption, \ac{WCET} of a task, and bandwidth requirements of messages  reflect a realistic scenario 
of current embedded \acp{MPSoC}.
We derived the energy consumption of each task on a certain \ac{PE} from the E3S benchmark and the communication energy consumption by a 
model proposed by~\cite{Hu:2003,Wolkotte:2005} with a link length of \SI{2}{\milli \meter} (resulting in  $E_{Lbit}=\SI{0.0936}{\nano \joule}$) and $E_{Sbit}=\SI{0.98}{\nano \joule}$   (see Section~\ref{sec:objectives}).

Furthermore, we selected a heterogeneous $6{\times}6$ \ac{NoC}-based architecture,consisting of three different processor types 
from~\cite{e3s}, including an IBM Power PC and variants of AMD K6.

\subsection{Considering Communication Constraints}

\begin{figure}
 \begin{tikzpicture} 

\begin{axis}[
    x tick label style={ /pgf/number format/1000 sep=}, 
		xtick = {0, 10, 20, 30, 40, 50, 60, 70, 80, 90,95},
    ylabel = {success rate in \%},
    xlabel = {utilization classes},
    width=\columnwidth,
    height=5.0cm,
    ymin = 0,
    xmax = 100,
    legend style={at={(0.80,0.9)}, anchor=north,legend columns=1},    
]

\addplot+[fill=gray!30, no marks, draw=none] coordinates {
	(0, 100)
	(10, 96.2962962962962)
	(20, 84.0579710144927)
	(30, 56.7708333333333)
	(40, 39.2575928008998)
	(50, 31.8385650224215)
	(60, 26.0579064587973)
	(70, 18.5085354896675)
	(80, 13.1313131313131)
	(85, 10.5882352941176)
	(90, 05.55555555555555)
	(95, 03.57142857142857)
	(0, 05)
    } \closedcycle;
		
\addplot[color=blue, mark=triangle*] coordinates {
	(0, 100)
	(10, 96.2962962962962)
	(20, 84.0579710144927)
	(30, 56.7708333333333)
	(40, 39.2575928008998)
	(50, 31.8385650224215)
	(60, 26.0579064587973)
	(70, 18.5085354896675)
	(80, 13.1313131313131)
	(85, 10.5882352941176)
	(90, 05.55555555555555)
	(95, 03.57142857142857)
    };

\addplot+[fill=white, no marks, draw=none] coordinates {
	(0, 04.8)
	(95, 03.57142857142857)
	(90, 05.55555555555555)
	(85, 05.88235294117647)
	(80, 07.07070707070707)
	(70, 06.28930817610062)
	(60, 06.57015590200445)
	(50, 07.17488789237668)
	(40, 13.1608548931383)
	(30, 25)
	(20, 64.4927536231884)
	(10, 87.037037037037)
	(0, 100)
	} \closedcycle;
	
	\addplot[color=red, mark=*] coordinates {
	
	(95, 03.57142857142857)
	(90, 05.55555555555555)
	(85, 05.88235294117647)
	(80, 07.07070707070707)
	(70, 06.28930817610062)
	(60, 06.57015590200445)
	(50, 07.17488789237668)
	(40, 13.1608548931383)
	(30, 25.)
	(20, 64.4927536231884)
	(10, 87.037037037037)
	(0, 100)
	};

\legend{,resource availability,, constraint solver} 

\end{axis} 

\end{tikzpicture}
 \caption{\label{fig:success_5x5}Success ratio of mapping operating points obtained for the E3S benchmarks to a 5$\times$5 \ac{NoC} for different 
utilization classes. 
Success ratios are given for resource management based on resource availability and resource management using a constraint solver are 
compared~\cite{WGWGT14}.}
\end{figure}
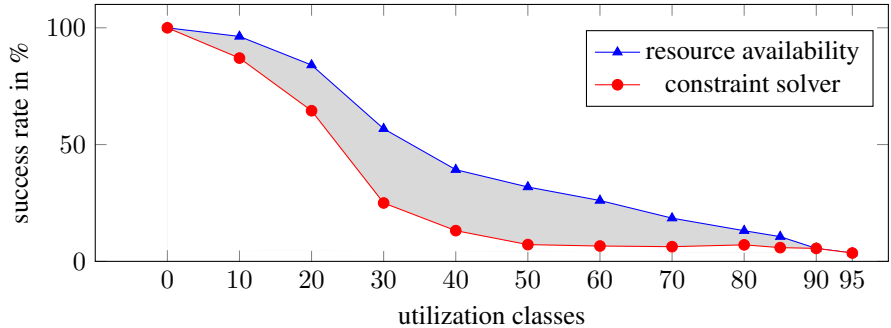

In a first experiment, we evaluate the influence of the communication constraints, i.e. \ref{const:hop} - \ref{const:sl}, on 
finding feasible mappings. {
As exemplified in Fig.~\ref{fig:mapping}, checking only the availability of the needed processing resources, e.g. as proposed in \cite{Ykman-Couvreur:2006,Shojaei:2013,WGT:2014} or assuming only point-to-point connections~\cite{Singh:2013b},  is not sufficient for a feasible mapping in a packet-switched \ac{NoC} architecture. }
Indeed, it only satisfies  \ref{const:type} and neglects the other constraints.
To visualize this, we tried, in 6,000 test cases, to map operating points from the above mentioned E3S benchmark applications to a 
preoccupied system using Algo.~\ref{algo:backtracking} without a timeout.
As a result, Fig.~\ref{fig:success_5x5} shows the gap between only considering the resource availability (blue curve) and the actual 
feasibility considering the communication constraints \ref{const:hop} and \ref{const:sl}  tested by the introduced constraint solver (red curve).
The utilization classes on the x-axis denote the percentage of utilized computing resources before testing to add the new application.
For example, 0 represents a completely empty system and the utilization class 10 includes systems where 1 to  \SI{10}{\percent} of the \acp{PE} are 
utilized by previously mapped applications.
The gray area between the two curves highlights the optimism introduced by a run-time system which only relies on computing resource 
availability as in~\cite{Singh:2013b}.
In case of a \SI{40}{\percent} utilization class, \SI{39}{\percent} of applications could be mapped to the system by only considering resource availability, 
while only for  \SI{13}{\percent} guarantees for holding their deadlines could be given.
All remaining ones miss deadlines because of communication latencies or are actually not  mapped because of congested communication resources. 
Overall, this underlines the importance of considering communication and routing constraints when it comes to methodologies for application 
mappings on composable \ac{NoC}-based \acp{MPSoC} with a predictable execution times.

\subsection{Temporal Isolation versus Spatial Isolation}

By applying the \ac{EA}-based \ac{DSE} illustrated in Fig.~\ref{fig:flowchart}, we generated and evaluated an overall of 200,000 mappings per application, resulting from a population size of 200 and 1000 iterations. 
For each of these mappings, we conducted the performance analysis proposed in Section~\ref{sec:predictability}.
This was done with the number of additional tasks set to $K_{max}{-}|\taskcluster|\,{=}\,4$, $\Snt\,{=}\SI{50}{\micro \second}$, and $\SIos\,{=}\SI{10}{\micro \second}$. 
As outlined in Section~\ref{sec:objectives}, the optimization criteria were minimizing (I)  the energy consumption of each mapping, (II) the number of routed messages, (V) the number of allocated PEs per resource type $r \in \RES$.
Further criteria were maximizing (III) the average and (IV) the minimal hop distance in order to generate more flexible mappings (the bigger 
the hop count of a message cluster, the less stringent becomes constraint~\ref{const:hops}).
Out of these 200,000 mappings, all Pareto-optimal solutions which do not violate the application deadline are stored as operating points together with the created constraint graphs and the values of the evaluated objectives (less than 100 points per application).

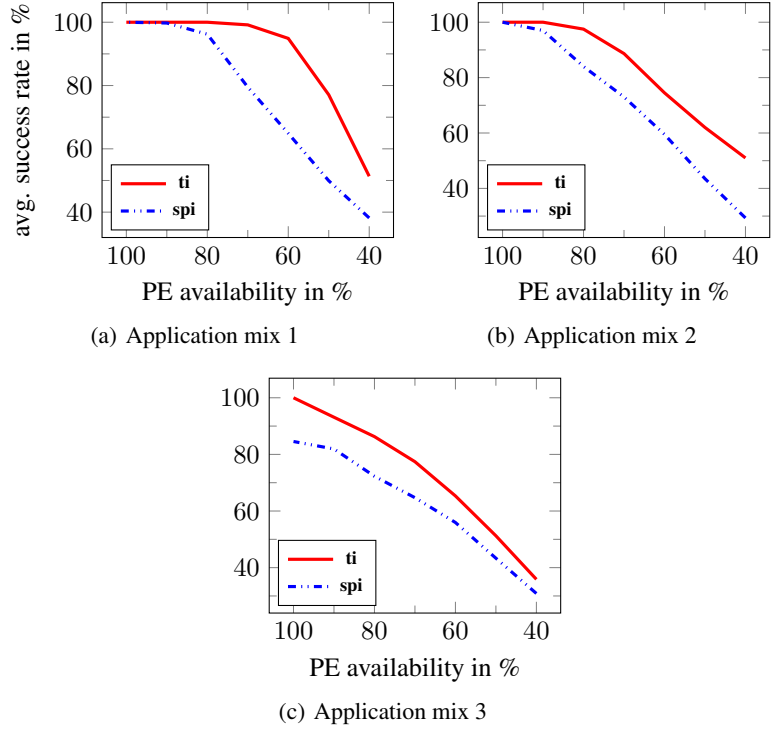
\begin{figure*}[ht!]
\begin{center}
\subfigure[Application mix 1]{
\pgfplotstableread{results/ReliabilityExperimentPareto.dat}{\pistonkinetics}
\begin{tikzpicture}
\begin{axis}[minor tick num=1,width = 0.45\linewidth,
xlabel=\ac{PE} availability in \%, ylabel = avg. success rate in \%,  legend style={legend pos=south west},x dir=reverse
]
\addplot [red,very thick] table [x expr=\thisrowno{0}*100, y expr=\thisrowno{2}*100] {\pistonkinetics};
\addplot [dashdotdotted,blue,very thick] table [x expr=\thisrowno{0}*100, y expr=\thisrowno{3}*100] {\pistonkinetics}; %
\legend{\scriptsize \textbf{ti}, \scriptsize \textbf{spi}}
\end{axis}
\end{tikzpicture}
\label{fig:rel1}
}
\subfigure[Application mix 2]{
\pgfplotstableread{results/ReliabilityExperimentPareto2_final.dat}{\pistonkinetics}
\begin{tikzpicture}
\begin{axis}[width = 0.45\linewidth, minor tick num=1,
xlabel=\ac{PE} availability in \%, 
legend style={legend pos=south west},x dir=reverse
]
\addplot [red,very thick] table [x expr=\thisrowno{0}*100,y expr=\thisrowno{2}*100] {\pistonkinetics};
\addplot [dashdotdotted,blue,very thick] table [x expr=\thisrowno{0}*100,  y expr=\thisrowno{3}*100] {\pistonkinetics};
\legend{\scriptsize \textbf{ti}, \scriptsize \textbf{spi}} 
\end{axis}
\end{tikzpicture}
\label{fig:rel2}}

\subfigure[Application mix 3]{
\pgfplotstableread{results/ReliabilityExperiment4Pareto2.dat}{\pistonkinetics}
\begin{tikzpicture}
\begin{axis}[width = 0.45\linewidth, minor tick num=1,
xlabel=\ac{PE} availability in \%, 
legend style={legend pos=south west},x dir=reverse
]
\addplot [red,very thick] table [x expr=\thisrowno{0}*100, y expr=\thisrowno{2}*100] {\pistonkinetics};
\addplot [dashdotdotted,blue,very thick] table [x expr=\thisrowno{0}*100, y expr=\thisrowno{3}*100] {\pistonkinetics};
\legend{\scriptsize \textbf{ti}, \scriptsize \textbf{spi}} 
\end{axis}
\end{tikzpicture}
\label{fig:rel3}
}
\caption{Evaluation of the average success rate of run-time mapping of pre-explored Pareto-optimal operating points belonging to different application mixes for 
spatial isolation (\textbf{spi}) and temporal isolation (\textbf{ti}) depending on the percentage of initially available \acp{PE}. 
The average success rate refers to the number of 
applications which could be successfully mapped in an overall of 100 experiments, providing a good measure for the system utilization.}
\end{center}
\end{figure*}

We then implemented an \ac{RM} for mapping different run-time mixes of the benchmark applications, where the applications are mapped iteratively. 
The operating points of each application were sorted in increasing order of energy consumption values (the objective of 
main interest in our experiments). 
In this order, a \emph{run-time embedder}, following a first-fit scheme, searches the first operating point whose constraint graph can be 
feasibly mapped to the system. 
For comparison, we implemented two embedder variants based on Algo.~\ref{algo:backtracking}: 
(a) variant \textbf{ti} performs the proposed mapping with \emph{temporal isolation} and (b) variant \textbf{spi} with \emph{spatial 
isolation} 
(see \cite{WGWGT14}).\footnote{In both variants, only operating points are used which do not violate the deadline, hence both satisfy the 
real-time requirements.}
{
These embedders try to map one constraint graph of each application (from the first fitting \ac{OP})  to the architecture.
Here, the mapping of the applications is incremental, i.e.  first, a constraint graph from the first application is mapped, then a  constraint graph  from the second application etc.
This simulates the arrival of different applications at different points in time during run time that constitute an application mix, which was unknown at design time.
In principle, the proposed run-time mapping would also support the remapping of \acp{OP} and removing of mapped applications but this is not considered in the following experiments.
}

We evaluated how many applications out of an application mix we can map successfully to our system (referred as \emph{success rate} in the following) for both variants.
For three different application mixes, experiments were repeatedly performed, but \acp{PE} were successively made unavailable for mapping 
any tasks so that the overall \ac{PE} availability ranged from $100\%$ down to $40\%$ (which also captures scenarios with, e.g., faulty or 
powered down \acp{PE}). 
We generated $100$ different sequences in which \acp{PE} are randomly made unavailable, starting from $100\%$ availability of 
\acp{PE} down to  $40\%$, and used the average values per number of available \acp{PE} as the result.

%

The result of such a set of experiments is depicted in Fig.~\ref{fig:rel1} for \textit{application mix~1} consisting of one telecom application and two networking applications. 
\textit{Application mix~2} (see Fig.~\ref{fig:rel2}) is composed of one telecom, three automotive and one consumer application, while \textit{application mix~3} (see Fig.~\ref{fig:rel3}) consists of two automotive, two consumer and two networking applications. 
In the graphs, the x-axis represents the percentage of initially  available \acp{PE} while the y-axis corresponds to the ratio of successful mappings.
The main trend observed is that with decreasing \ac{PE} availability, the success rate declines much faster when using spatial isolation.
In the case of \textit{application mix~1}, the success rate of \textbf{spi} drops to $65\%$ while it still remains at $95\%$
using the proposed \textbf{ti} in case of an availability of $60\%$ of the \acp{PE}.
The experiments with \textit{application mix~2} show a similar behavior. 
Even more drastically, in the experiments with \textit{application mix~3}, all applications could be mapped with our proposed approach in
the case where all
\acp{PE} are available, whereas using \textbf{spi}, one application in the mix could not even be mapped at all.
 
In our test cases, the obtained energy consumptions of \textbf{ti} mappings were always equal or better than those using
\textbf{spi} mappings for a \ac{PE} availability of $100\%$ for \textit{application mixes~1} and \textit{2}. 
In \textit{application mix~1}, \textbf{ti} and \textbf{spi} reached the same results. 
In \textit{application mix~2}, \textbf{ti} mapped operating points with an energy consumption of \SI{351}{\milli \joule}, whereas \textbf{spi}
mapping resulted in \SI{477}{\milli \joule} per execution. 
Being able to obtain run-time application mappings which are better with respect to the objective (energy) is a direct 
consequence of being able to better utilize the available resources.
For all other rates of \ac{PE} availability and also for $100\%$ \ac{PE} availability in \textit{application mix~3}, a comparison is not 
meaningful
as \textbf{spi} is not able to map as many applications as \textbf{ti}.


\subsection{Execution Time}

\begin{figure*}[t]
	\centering
	\subfigure[Execution times successful mappings.]{
\begin{tikzpicture}
\pgfplotstableread{results/executiontime_cdf_success_300mhz.dat}{\pistonkinetics}
\begin{axis}[minor tick num=1,
width = 0.47\linewidth,
height = 6 cm,
xlabel=execution time in ms ,
ymin = 0,
ymax = 105,
ylabel = {CDF} in  \%, 
legend style={legend pos=south east},
 ytick={0,25,50,75,97},
grid=both,
minor grid style={dotted},
legend style={legend pos=south east},
xmode = log,  
]

\addplot [red,very thick] table [x expr=\thisrowno{0}, y expr=\thisrowno{1}*100] {\pistonkinetics};

\draw[draw=black, densely dashed] (axis cs: 305 ,0) -- node[sloped, above] {305\,ms} (axis cs: 305 ,110);
\end{axis}
\end{tikzpicture}
\label{fig:execsuc}
}
\subfigure[Execution times of failed mappings.]{
\begin{tikzpicture}
\pgfplotstableread{results/executiontime_cdf_fail_300mhz.dat}{\pistonkinetics}
\begin{axis}[minor tick num=1,
width = 0.49\linewidth,
height = 6 cm,
xlabel=execution time in ms ,
ymin = 0,
ymax = 105,
ylabel = {CDF} in  \%, 
legend style={legend pos=south east},
grid=both,
 ytick={0,25,50,75,95},
minor ytick={100},
grid=both,
minor grid style={dotted},
xmode = log,  
]

\addplot [red,very thick] table [x expr=\thisrowno{0}, y expr=\thisrowno{1}*100] {\pistonkinetics};

\draw[draw=black, densely dashed] (axis cs: 305 ,0) -- node[sloped, above] {305\,ms} (axis cs: 305 ,110);
\end{axis}
\end{tikzpicture}
\label{fig:execfail}
}
\caption{
CDF of backtracking algorithm execution times on a $32$ bit embedded processor at \SI{300}{\mega \hertz}.  
Note the logarithmic scale of the x-axis in both plots. 
The maximal needed execution time for finding a feasible mapping (\SI{305}{\milli \second}) is presented by a dashed vertical line in both plots. }
\label{fig:exectimes}
\end{figure*}
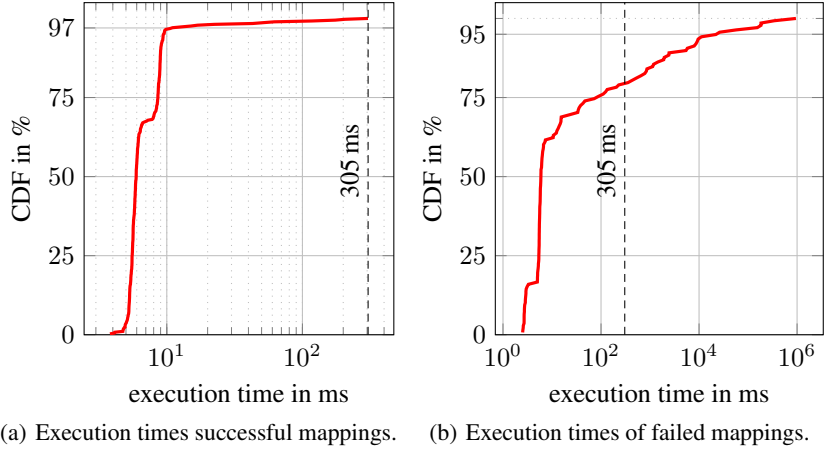

\begin{revision2}
 Constraint solving is the central concept for making use of the offline explored operating points at run time. 
 However, this implies an additional overhead for determining a feasible mapping based on the provided constraint graphs. 
 In this experiment, we evaluate the execution times of the run-time backtracking mapping algorithm (Algo.~\ref{algo:backtracking}) performed by a central \ac{RM}. 
 Here, we applied \cite{Roloff:2015} to simulate the execution of the \ac{RM} according to~\cite{WildermannWT15} on a \SI{32}{\bit} embedded processor with a clock frequency of \SI{300}{\mega \hertz}. 
 Overall, feasible mappings for $500$ constraint graphs on an $8 \times 8$ \ac{NoC} architecture were searched via the backtracking mapping algorithm. 
Figure~\ref{fig:exectimes} shows the \ac{CDF} of the execution times (in \si{\milli \second}) measured for executing the run-time backtracking algorithm.
 \end{revision2}
The \ac{CDF} describes the maximal execution time needed by the percentage of runs.
Values are separated for the cases of (a) successful (i.e. at least one feasible mapping exists) and (b) failed constraint solving (no feasible mapping exists). 
Note that constraint solving is a complex task (in the worst-case, Algo.~\ref{algo:backtracking} has exponential run time) and took up to 
\begin{revision2}
\SI{305}{\milli \second} (denoted in the Fig.~\ref{fig:exectimes} by a vertical line) for successful and \SI{947878}{\milli \second} for failed mappings. 
The vast majority of the applications can be mapped much faster, e.g., \SI{97}{\percent} of the successful test cases took at most \SI{10}{\milli \second}.  
In the case of failed mappings, execution times were much higher. 
Only \SI{78}{\percent}  of test cases took below \SI{305}{\milli \second}, and \SI{19}{\percent} took seconds or even minutes (see Fig.~\ref{fig:execfail}).
\end{revision2}

Note that this time only elapses before a newly arriving real-time application is started.
While we are dealing with applications that---once mapped---are periodically executed for a long time, mapping times in the range of few seconds might be tolerable. 
However, in order to bound the execution time of the run-time mapping and supporting domains where mapping time matters, we propose the usage of a timeout mechanism (see Algo.~\ref{algo:backtracking}): 
\begin{revision2}
We stop the algorithm after the expiration of the timeout interval and classify the currently tested mapping as infeasible.
\end{revision2}
The timeout value needs to be appropriately chosen to fulfill the turn-around time requirements of the application being mapped.
Particularly, as a too low value may increase the number of false negatives (i.e. feasible mappings which are classified as infeasible).
\begin{revision2}
 However, for our experiments even with a timeout value as low as \SI{10}{\milli \second}, we would only reject feasible mappings (i.e., classify false negatives) in \SI{3}{\percent} of the cases. 
 As we provide multiple operating points per application, a mapping according to another constraint graph may then be obtained.
\end{revision2}

{
Nevertheless,  to handle larger systems  the execution times of this algorithm may be not acceptable anymore.
Therefore,  we will conduct further research on the run-time \ac{CSP} solving.
This may include a hierarchical decomposition of the architecture where the backtracking algorithm searches in a sub-architecture first, distributed \ac{CSP} solving,  or dedicated hardware support~\cite{whz15}. 
}
\begin{revision2}
With using isolated regions per applications, also fast heuristics solving a 2D packing problem can be used~\cite{WeichslgartnerW16}. 
However, this makes use of spatial isolation only and makes temporal isolation infeasible, thus, decreasing the utilization.
\end{revision2}

%

\section{Conclusions}
\label{sec:conclusion}
In this article, we proposed a technique to increase the utilization of many-core systems using hybrid application mapping combined with a static performance analysis considering bounds on temporal interference on tasks.
More specifically, the design-time analysis for applications with real-time constraints was performed considering, for the first time in a hybrid application mapping approach, temporal isolation of concurrent tasks with bounds on task interference.
Via  \acf{DSE} of mappings, a set of Pareto-optimal operating points with composable performance values is obtained.
The subsequent operating point mapping at run time is achieved by solving a constraint satisfaction problem.
It has been shown that this hybrid approach allows to provide predictable application mappings within high system utilization and reduced  number of \acp{PE} that are needed to execute various application mixes while satisfying real-time requirements.
Another major advantage of our approach over previous work is the reduction of the exploitative search for feasible mappings to design time and leave only the remaining freedom in finding a concrete mapping to the \acf{RM}.
This was possible through the concept of a constraint graph characterizing feasible mappings.
In the future, we want to investigate scalability enhancements of our run-time mapping approach, e.g. distributed constraint solving techniques.

\section*{Acknowledgment}
This work was supported by the German Research Foundation
(DFG) as part of the Transregional Collaborative Research
Center “Invasive Computing” (SFB/TR 89).
\bibliographystyle{ACM-Reference-Format-Journals}
\bibliography{literature}



\medskip


%

%

\end{document}